\documentclass{article}
\usepackage[margin=2.5cm]{geometry}
\usepackage{xcolor}
\usepackage[hidelinks]{hyperref}
\usepackage{bm}
\usepackage{graphicx}
\usepackage{subcaption}
\usepackage{float}
\usepackage{amsmath,amssymb,siunitx}
\usepackage{algorithm2e}
\usepackage{tikz}
\usepackage{needspace}
\usetikzlibrary{arrows.meta,calc,positioning}

\definecolor{phasebackground}{RGB}{176,196,222}

\title{$N$-phase flow simulations using X-Mesh and an explicit marker interface representation}
\author{
Antoine Quiriny\thanks{UCLouvain, Institute of Mechanics, Materials and Civil Engineering (iMMC), Louvain-la-Neuve, Belgium}\kern0.4em\thanks{Corresponding author. Email: \href{mailto:antoine.quiriny@uclouvain.be}{antoine.quiriny@uclouvain.be}}
\and
Jonathan Lambrechts\footnotemark[1]
\and
Michel Henry\thanks{Univ. Grenoble Alpes, Inria, 38000 Grenoble, France. Email: \href{mailto:michel.henry@inria.fr}{michel.henry@inria.fr}}
\and
Nicolas Mo\"es\footnotemark[1]
\and
Jean-Fran\c{c}ois Remacle\footnotemark[1]
}
\date{}

\begin{document}

\maketitle
\begin{abstract}
We present a marker-based X-Mesh framework for incompressible $N$-phase flow simulations with sharp, explicitly tracked interfaces. The interface is represented by Lagrangian marker chains, while the computational mesh is deformed so that selected mesh edges conform to the tracked geometry without changing mesh connectivity. This combination preserves the advantages of body-fitted finite element discretizations while allowing highly distorted and nearly degenerated elements to capture fine interfacial features on relatively coarse meshes without requiring mesh refinement. The arbitrary Lagrangian--Eulerian (ALE) finite element solver uses a pressure field that is continuous within each phase and discontinuous across the fitted interface. Surface tension is introduced through a variational membrane, so the pressure jump is resolved without prescribing a curvature-dependent correction; a Young wall energy provides the corresponding treatment of fluid--fluid--solid triple points. Topological changes are handled geometrically from the marker representation. Numerical experiments illustrate the ability of the approach to transport complex interfaces and to control topology changes through marker-based feature-size criteria.
\end{abstract}

\noindent\textbf{Keywords:} Multiphase flow; X-Mesh; Front tracking; Finite element method; Topology change; Surface tension.

\section{Introduction}
\label{sec:intro}
$N$-phase flow simulations play a central role in numerous industrial and scientific applications, including atomization processes, inkjet printing, additive manufacturing, microfluidics, boiling phenomena, and multiphase flows encountered in energy production systems. In such configurations, the accurate prediction of interface dynamics is essential since capillary effects, topology changes, and interfacial stress discontinuities strongly influence the global flow behavior. The numerical simulation of $N$-phase flows therefore requires methods capable of accurately tracking evolving interfaces while maintaining robustness in the presence of large deformations and complex geometrical configurations.

Among the numerical approaches developed for incompressible two-phase flows, finite element methods constitute an attractive framework due to their geometric flexibility, their natural treatment of complex boundaries, and their ability to accurately approximate interfacial jump conditions. Over the past decades, numerous finite element formulations have been proposed in combination with interface-capturing or interface-tracking strategies, including Arbitrary Lagrangian-Eulerian (ALE) methods \cite{Hirt1974ALE,Donea1982ALEFE}, level-set methods \cite{OsherSethian1988LevelSet,Sussman1994LevelSetTwoPhase}, Volume-of-Fluid techniques \cite{HirtNichols1981VOF}, phase-field approaches \cite{Jacqmin1999PhaseField}, and unfitted formulations such as XFEM \cite{Moes1999XFEM} or CutFEM \cite{Burman2015CutFEM}. While these approaches have demonstrated significant success, the simultaneous achievement of sharp interface representation, robustness under large interface deformations, and efficient handling of topology changes remains a challenging issue.

Front-tracking methods are a classical approach to the numerical simulation of sharp-interface multiphase flows, in which the interface is represented by a set of Lagrangian markers advected by the flow velocity field. This class of methods has been widely investigated in the literature, notably in the pioneering works of Unverdi and Tryggvason \cite{unverdi1992front}, Tryggvason et al. \cite{tryggvason2011direct}, and Popinet \cite{popinet2009overview,popinet_fronttracking}. Compared to implicit representations such as level-set methods, explicit marker descriptions provide a direct geometrical representation of the interface and allow a precise control of the local interface resolution. The counterpart of this explicit geometrical control is that topological changes are not handled automatically: merging, breakup, or contact events have to be detected and treated by dedicated algorithms. This requirement makes front-tracking methods more delicate to implement robustly, whereas level-set and interface-capturing methods were specifically introduced, in part, to avoid explicit connectivity management during topology changes.

Recently, the X-Mesh method introduced a new paradigm for moving interface problems by allowing arbitrarily large mesh deformations while preserving a fixed mesh connectivity \cite{Moes2023XMeshStefan}. The key idea of the method is to explicitly conform the mesh to the interface while authorizing the degeneration of mesh elements whenever necessary. As a consequence, the method avoids remeshing operations even in the presence of extreme interface deformations and topological transitions. In the context of two-phase flows, this approach enables a sharp representation of the interface together with an accurate treatment of interfacial discontinuities, while preserving the advantages of conforming finite element discretizations \cite{Quiriny2024XMeshTwoPhase}.

Other X-Mesh formulations for two-phase flows relied on a level-set representation of the interface \cite{Quiriny2024XMeshTwoPhase}. Although this approach provides a convenient implicit description of evolving geometries, the interface resolution naturally remains tied to the local mesh size. In particular, thin structures such as liquid filaments become difficult to represent once their characteristic thickness approaches the mesh resolution. However, accurately capturing the dynamics of subgrid-scale filaments before topological transitions may be essential in many applications involving breakup or coalescence phenomena.

The present work addresses this limitation by replacing the implicit level-set description with an explicit Lagrangian marker representation of the interface. The interface is described by connected marker particles forming a front-tracking representation that evolves according to the local fluid velocity. This explicit geometric description enables a direct control of the interface resolution independently of the background mesh size. In particular, thin filaments smaller than the characteristic mesh size may still be represented by introducing arbitrarily degenerated mesh elements inside these structures. Such a strategy naturally fits within the X-Mesh framework since the method was specifically designed to accommodate highly distorted and vanishing elements without modifying mesh connectivity.

The method also handles topological changes: after marker advection, intersecting interfaces are reconstructed into a consistent phase partition, allowing coalescence and breakup. A separate feature-size criterion controls the removal of thin films and filaments. Both operations modify the marker-interface connectivity while preserving the connectivity of the computational mesh.

The marker representation also provides a convenient way to handle zero-dimensional geometric features in two-dimensional simulations. These features are explicitly tracked at the marker level and can therefore be preserved during the mesh deformation process by aligning selected mesh nodes with their positions. In two-phase flow, an important example is a fluid--fluid--solid triple point. In the present formulation, such contact points are coupled to a variational membrane model for surface tension and to a Young wall energy, so that they may move along an impermeable wall toward a prescribed equilibrium contact angle. More generally, the same geometric capability may be useful beyond two-phase flows, for example in fluid--structure interaction problems where sharp corners or localized boundary features have to be accurately represented on the computational mesh.

The objective of the present work is therefore to develop a marker-based X-Mesh framework for incompressible $N$-phase flows combining %
Lagrangian front tracking, conforming finite element discretizations, and topology-aware mesh deformations without remeshing. The fluid discretization uses a pressure space that is continuous within each phase and discontinuous across the fitted interface, while surface tension is represented by a variational membrane. The proposed formulation therefore resolves the pressure jump as part of the coupled problem, without prescribing a curvature-dependent correction, and extends the sharp-interface capabilities of previous X-Mesh approaches to highly localized interfacial structures and moving contact points. The formulation supports an arbitrary number $N$ of fluid phases. The test cases presented here involve at most three fluids; this restriction concerns only the examples, not the method.

The remainder of the paper is organized as follows. Section~\ref{sec:interface} introduces the explicit marker representation of the interface and the associated topological operations. This marker-tracking component is presented for completeness, since several alternative front-tracking strategies and implementations are available in the literature, although their algorithmic details are not always stated explicitly. Nothing in the proposed framework prevents the use of another marker-tracking approach or an existing implementation. The main focus of the paper is the mesh-deformation procedure described in Section~\ref{sec:xmesh}, which couples the marker interface with the X-Mesh framework. The $N$-phase flow formulation is then recalled in Section~\ref{sec:fluid}, and numerical experiments are reported in Section~\ref{sec:results}.

We conclude this introduction with an overview of the coupling between the objects used in the fluid simulation: the marker interface, the computational mesh and its phase labels, and the discrete velocity and pressure fields. The known velocity advects the markers; redistribution and topology control update the interface. X-Mesh then deforms the mesh at fixed connectivity, and the marker partition determines the element phases. Finally, the ALE finite element solver computes the new velocity and pressure on this updated geometry, including membrane and wall-energy forces. Section~\ref{sec:coupling_strategy} details this sequence.

\section{Explicit interface representation with markers}
\label{sec:interface}
%

In the present work, the interface $\Gamma$ is represented in two dimensions by a set of connected marker particles $\mathbf{x}_i$ linked by line segments. Let $\mathbf{u}(\mathbf{x},t)$ denote the fluid velocity, obtained from the incompressible Navier--Stokes problem in Section~\ref{sec:fluid}, or prescribed analytically in the advection test. The marker positions obey the kinematic advection equation:
\begin{equation}
    \label{advection_marker}
    \frac{d\mathbf{x}_i}{dt} = \mathbf{u}(\mathbf{x}_i,t).
\end{equation}

To improve the temporal accuracy of the interface advection, the marker trajectories are integrated using a second order Runge-Kutta discretization:
\begin{align}
    \mathbf{x}_i^{n+\frac{1}{2}}
    &=
    \mathbf{x}_i^n
    +
    \frac{\Delta t}{2} \,
    \mathbf{u}(\mathbf{x}_i^n) \\
    \mathbf{x}_i^{n+1}
    &=
    \mathbf{x}_i^n
    +
    \Delta t \,
    \mathbf{u}(\mathbf{x}_i^{n+\frac{1}{2}})   
\end{align}
In practice, the velocity at the marker positions is obtained through interpolation from the underlying fluid mesh.

Since the marker distribution directly controls the geometrical accuracy of the interface representation, maintaining a sufficiently regular spacing between markers is essential. If the distance between markers becomes too large, the interface curvature cannot be accurately approximated and fine geometrical structures may be lost. %
Conversely, the marker interface and the advecting velocity are represented on different discretizations. Marker spacing far below the fluid-mesh size may over-resolve elementwise velocity-gradient jumps and produce spurious interface oscillations \cite{PopinetZaleski1999FrontTracking}.

\begin{figure}[htbp]
    \centering
    \includegraphics[width=0.95\textwidth]{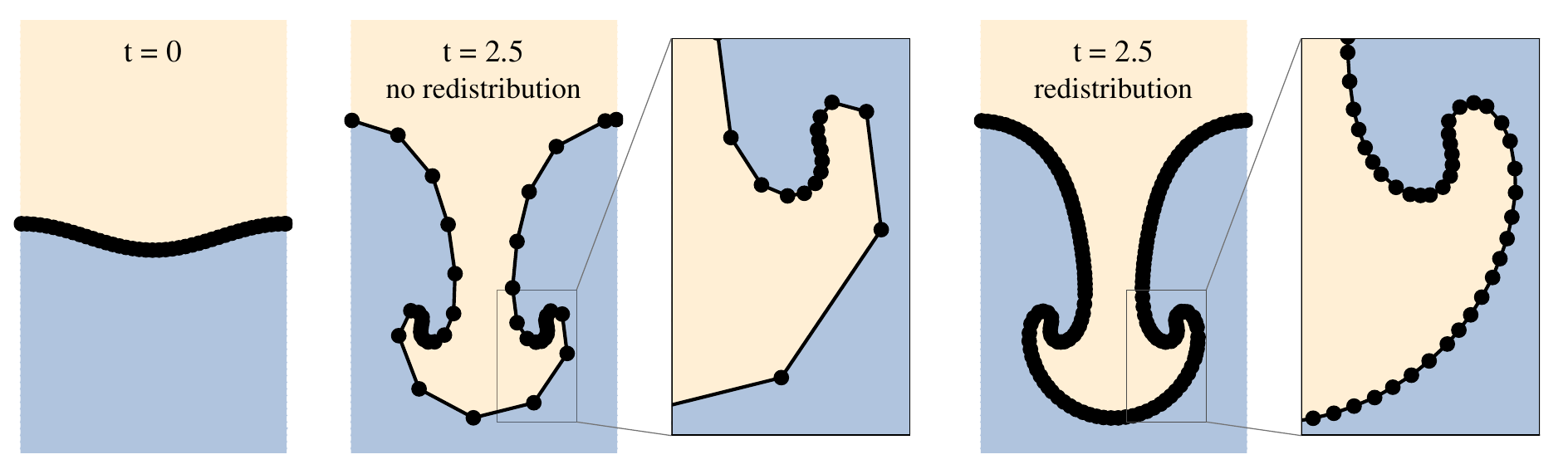}
    \caption{Effect of the redistribution procedure on the marker spacing along a Rayleigh-Taylor interface. Initial marker positions (left), final marker positions without redistribution (center), and final marker positions with redistribution (right). }
    \label{fig:redist_markers}
\end{figure}

To avoid these issues, a redistribution procedure is periodically applied along the interface. Markers separated by a distance smaller than a prescribed threshold are merged, while additional markers are inserted in regions where the local spacing exceeds a maximal admissible distance. This redistribution procedure maintains an approximately uniform arc-length discretization of the interface while preserving its geometrical features. An example of this redistribution process is shown in Figure~\ref{fig:redist_markers} for a Rayleigh-Taylor instability, where the marker spacing is regularized along the increasingly deformed interface. At impermeable walls, contact points move tangentially (Section~\ref{sec:boundary_markers}). Artificial feature-size offsets are clipped to the domain, whereas numerical overshoots during advection require a smaller time step.

\subsection{Oriented phase-partition representation}

Nested, disjoint closed curves can be organized by an inclusion tree. Here we use an oriented marker graph so that open chains and shared junction nodes can be represented explicitly as well. This choice simplifies the bookkeeping by storing segment orientations and adjacent phases in a single structure. Each ordered fluid-interface
segment $e=(a,b)$ carries the two physical phases adjacent to it:
$\phi_e^L$ on its left and $\phi_e^R$ on its right. At a junction, all
incident interface chains share the same marker node, while the phase
information remains attached to the oriented segments.

This representation, referred to as the phase topology, contains the shared
marker nodes, the oriented interface edges, and the connected regions occupied
by each physical phase. The boundary of a phase region is stored as one or
several closed cycles of oriented half-edges. Every fluid-interface edge is
therefore referenced exactly twice: once by the region on its left and once,
with the opposite orientation, by the region on its right. A physical phase
may occupy several disconnected regions, and the set of phases incident to a
shared node identifies triple points and more general multiphase junctions.

\begin{figure}[htbp]
    \centering
    \includegraphics[width=0.46\textwidth]{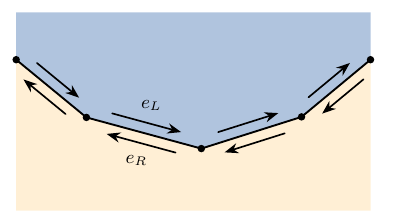}
    \caption{Four consecutive edges from the redistributed interface in Figure~\ref{fig:redist_markers}. Each edge has two oppositely oriented half-edges, shown offset for clarity.}
    \label{fig:halfedges}
\end{figure}

Figure~\ref{fig:halfedges} illustrates the two orientations along a discrete interface.

Before the marker motion, the valid non-intersecting configuration is converted
into such an oriented partition, or the existing partition is reused when it
still matches the marker graph. Each previous phase region stores its physical
phase, its oriented boundary cycles, and its geometrical depth. During marker
advection, the marker positions move while the oriented edges and their
left/right phase labels provide the physical memory needed to classify the
regions created by a subsequent topological event.

\subsection{Treatment of topological changes}

One of the main advantages of the marker representation is that topological events can be detected directly from the geometry of the transported front. The method used here does not prescribe a reconnection rule between marker chains. Instead, it reconstructs the phase partition after advection and then extracts the new interfaces as the boundaries between phases.

The procedure starts from a valid set of non-self-intersecting interfaces and
from the associated oriented phase partition. The markers are first advected
with the fluid velocity. If this motion produces intersections between
interface segments, the intersecting segments are split at the intersection
points and the resulting planar graph is used as the input geometry for the
reconstruction of the interfaces.\\
The input state is shown first, followed by six reconstruction steps.\\
\begin{minipage}{0.5\textwidth}
\textbf{Input. Previous oriented phase partition}\\
We start from a valid, non-self-intersecting marker configuration and construct
the corresponding oriented phase partition. Its shared nodes,
left/right phase labels, connected phase regions, oriented boundary cycles,
and region depths form the physical reference state for the reconstruction.
This representation is transported with the markers even when the moved
interface graph is no longer a valid non-intersecting partition.
\end{minipage}%
\hfill
\begin{minipage}{0.45\textwidth}
\includegraphics[width=\linewidth,trim={0 1.75cm 0 0},clip]{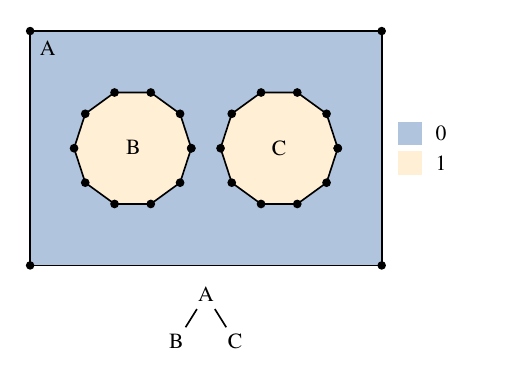}
\end{minipage}\\
\begin{minipage}{0.5\textwidth}
\textbf{Step 1. Move the markers}\\
The markers are advected with the local fluid velocity, as described in equation \ref{advection_marker}. This Lagrangian update deforms each interface while preserving its marker-based representation. During this step, initially disjoint interfaces may approach each other, overlap, or self-intersect, thereby creating the geometric configurations from which topological changes, such as coalescence or breakup, must be detected. An example of such a merging configuration is shown in the right panel.
\end{minipage}%
\hfill
\begin{minipage}{0.45\textwidth}
\includegraphics[width=\linewidth]{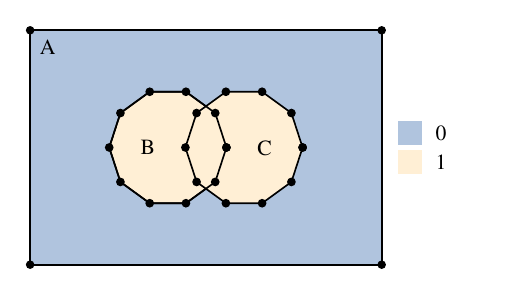}
\end{minipage}%
\\
\begin{minipage}{0.5\textwidth}
\textbf{Step 2. Intersect the interfaces}\\
After the marker advection, all intersections between interface segments are identified. This includes intersections between distinct interfaces as well as self-intersections of a single interface. To avoid a quadratic search over all segment pairs, the detection is performed using a spatial partitioning structure that selects only nearby candidate segments. The resulting intersection points are then inserted into the marker set, so that the subsequent reconstruction is performed on a conforming set of interface fragments.
\end{minipage}%
\hfill
\begin{minipage}{0.45\textwidth}
\includegraphics[width=\linewidth]{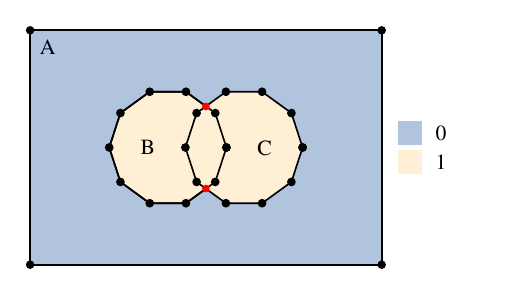}
\end{minipage}%
\\
\begin{minipage}{0.5\textwidth}
\textbf{Step 3. Triangulate the markers}\\
The updated marker set is then triangulated by imposing the interface fragments as constrained edges of the triangulation. This produces a planar subdivision that is conformal with respect to all detected intersections and interface segments. In practice, this constrained triangulation is performed using the algorithms available in Gmsh [\cite{geuzaine2009gmsh}], which provide a robust way to recover the prescribed segments while generating the auxiliary triangles required for the phase reconstruction.
\end{minipage}%
\hfill
\begin{minipage}{0.45\textwidth}
\includegraphics[width=\linewidth]{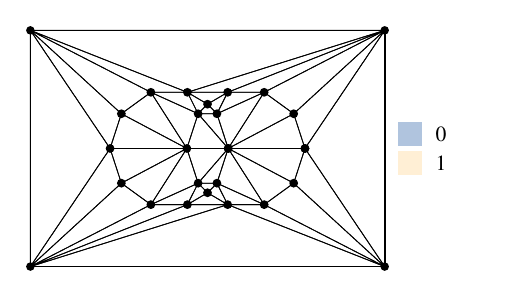}
\end{minipage}%
\\
\begin{minipage}{0.5\textwidth}
\textbf{Step 4. Winding number}\\
For each triangle of the constrained triangulation, the winding number
[\cite{HORMANN2001131}] of its barycenter is computed with respect to the
oriented boundary cycles of every previous phase region. This provides a local
topological signature indicating which previous regions contain the
triangle. Adjacent triangles carrying the same signature form a connected
preliminary component, which results in the colored partition shown in the
right panel. At this stage, these colors should only be interpreted as
topological labels; the corresponding physical phases are assigned in the
next step.
Here, `previous' refers to the phase labels and boundary connectivity recorded before the current advection step. The winding numbers use those cycles at the advected marker positions, not at their original coordinates.
\end{minipage}%
\hspace{0.045\textwidth}
\begin{minipage}{0.45\textwidth}
\includegraphics[width=\linewidth]{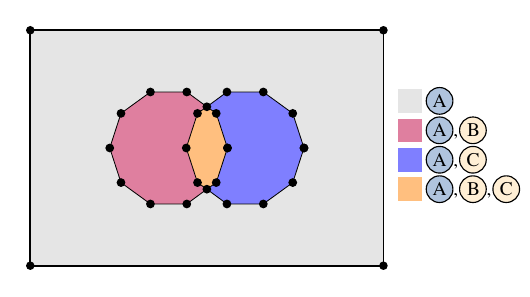}
\end{minipage}%
\\
\begin{minipage}{0.5\textwidth}
\textbf{Step 5. Choose phase}\\
Triangles with the same winding signature form a connected component. Its candidate phases are the transported regions in that signature. We select the candidate of greatest nesting depth for the whole component. Depth is the minimum number of interfaces separating a region from the exterior boundary, computed by BFS on the dual graph of phase regions. Exterior regions have depth zero. In this example, B and C have the same physical phase, so their overlap receives that phase as well.
\end{minipage}%
\hspace{0.045\textwidth}
\begin{minipage}{0.45\textwidth}
\includegraphics[width=\linewidth,trim={0 1.75cm 0 0},clip]{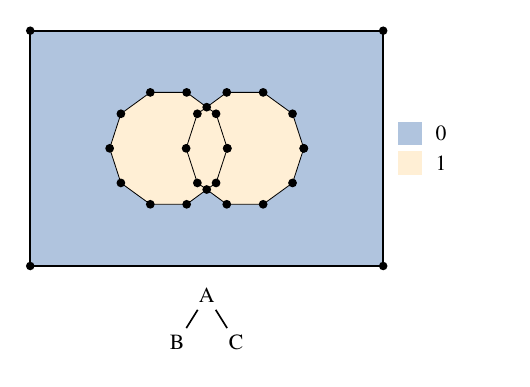}
\end{minipage} \\
\\
\begin{minipage}{0.5\textwidth}
\textbf{Step 6. Compute new interfaces}\\
Once a physical phase has been assigned to each triangle, the updated interfaces are extracted from the triangulation. More precisely, every edge shared by two triangles carrying different phases is identified as part of a new interface. These edges are then connected and ordered to form the reconstructed marker curves. The resulting interfaces define a valid, non-self-intersecting configuration that can be used as the input of the next time step.
\end{minipage}%
\hspace{0.045\textwidth}
\begin{minipage}{0.45\textwidth}
\includegraphics[width=\linewidth]{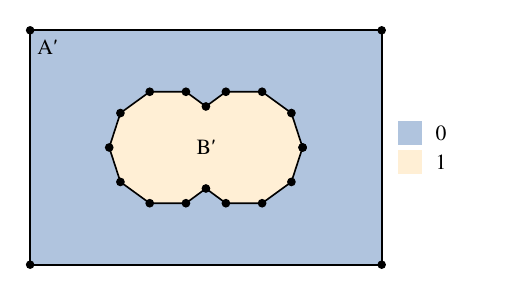}
\end{minipage}%

This reconstruction naturally handles both coalescence and breakup events. Instead of prescribing how individual marker chains should reconnect after a topological change, the method rebuilds the interface as the set of boundaries separating the reconstructed phases. The new marker chains are therefore obtained directly from the phase geometry, which provides a unified treatment of merging and splitting events.

The intersection search uses the interface geometry at the end of the advection step, rather than the surfaces swept by moving segments. If one filament passes completely through another within a single time step and the final segments do not intersect, this event can therefore be missed. Resolving such encounters requires a sufficiently small time step relative to the local separation and relative interface velocity. A continuous collision detector could provide an additional safeguard, but is not part of the present algorithm.

If different candidate phases have the same maximum depth, a prescribed phase-priority list breaks the tie. The selected phase is assigned to the whole connected signature component.

\Needspace{0.85\textheight}
\subsection{Why reconstruct the phase partition?}
\label{sec:nontrivial_topology}
\begin{figure}[H]
    \centering
    \includegraphics[width=0.96\textwidth]{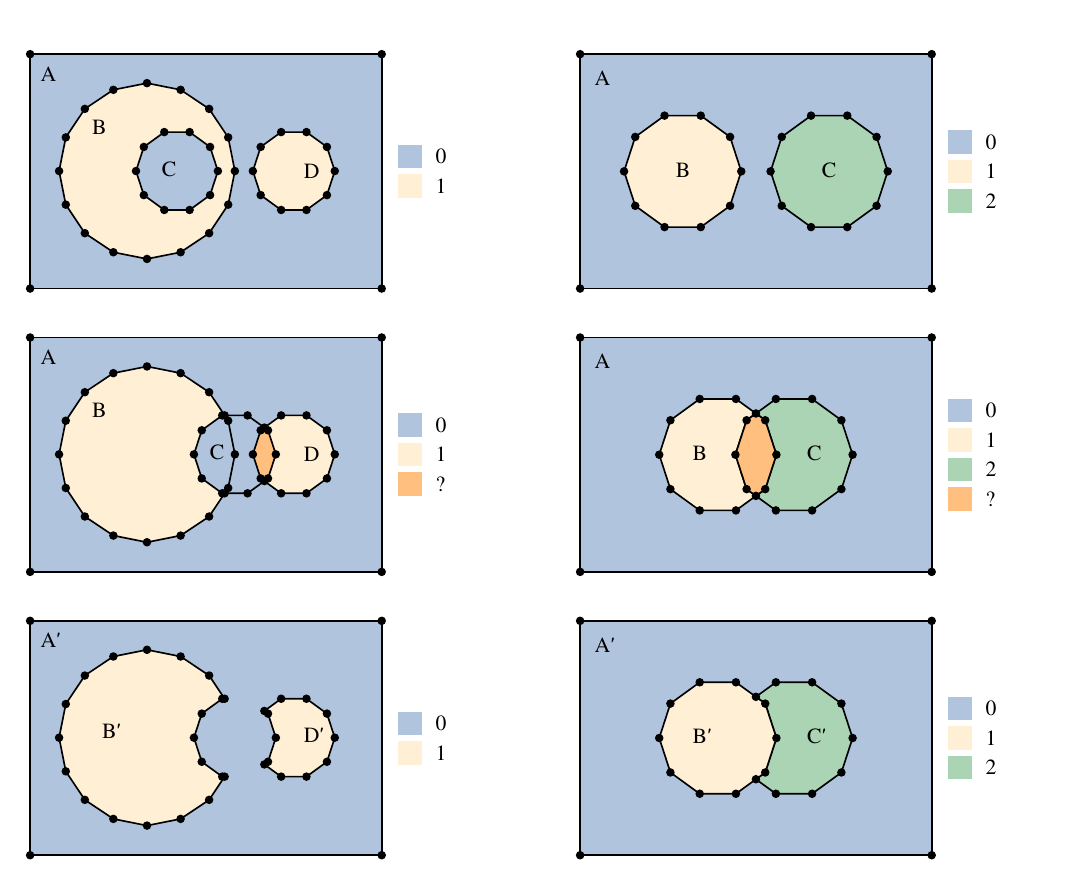}
    \caption{Two non-trivial reconstruction cases. Left: depth selects phase 0 in the overlap. Right: B and C have different phases. Read each column from the initial state to reconstruction. Black disks are markers; primes denote reconstructed regions. Phase labels appear on the right; orange (``?) marks overlaps awaiting phase selection. Schematic configurations.}
    \label{fig:nontrivial_topology}
\end{figure}

Intersecting marker chains do not by themselves define a unique phase partition. The crossing locations identify where the geometry must be split, but the transported region labels and their previous depths determine which phase each resulting component receives. Figure~\ref{fig:nontrivial_topology} illustrates two cases.

In the first case, A and C have phase 0, while B and D have phase 1. Initially, C lies inside B: their recorded depths are $d_A=0$, $d_B=d_D=1$ and $d_C=2$. After C moves across B and overlaps D, both C and D are candidates in their overlap. The inherited depth selects C, hence phase 0, although C is no longer enclosed by B. Recomputing a simple inclusion relation after motion would lose this information. Only boundaries separating different selected phases remain.

The second case reuses the geometry of the preceding step-by-step example, now with B in phase 1 and C in phase 2. Steps 1--4 identify the same overlap. Both candidates have depth one, so depth alone cannot decide. With the illustrative priority $1\succ2$, Step 5 assigns the entire overlap to phase 1. Step 6 retains an interface between phases 1 and 2, with two triple junctions; merging B and C as in the same-phase example would erase that interface. Reversing the priority would instead assign the overlap to phase 2.

\Needspace{4\baselineskip}
\subsection{Feature-size control through artificial topological changes}

As long as the velocity remains sufficiently regular, uniqueness of marker trajectories preserves the topology of a material interface. This property follows from regularity of the flow map, rather than from incompressibility alone. In practice, however, physical topological transitions are preceded by the formation of extremely thin liquid films or filaments whose thickness may become several orders of magnitude smaller than the characteristic flow scales.

As a film or filament thins, its thickness may fall below the local mesh size while the interface remains geometrically connected. The surrounding fluid discretization then cannot resolve the flow across that thickness, and the present model does not include a microscopic rupture law. As a consequence, numerical topological changes may occur prematurely due either to insufficient interface resolution or to the accumulation of discretization errors. Depending on the targeted application, preserving or removing these subgrid-scale structures may be desirable.

\begin{figure}[htbp]
    \centering
    \begin{tabular}{@{}c@{\hspace{0.06\textwidth}}c@{}}
    \includegraphics[width=0.42\textwidth]{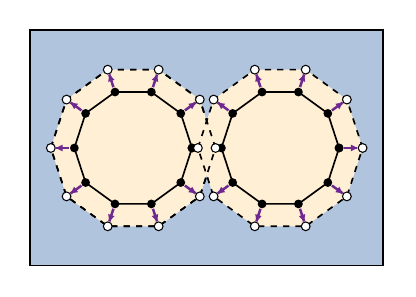} &
    \includegraphics[width=0.42\textwidth]{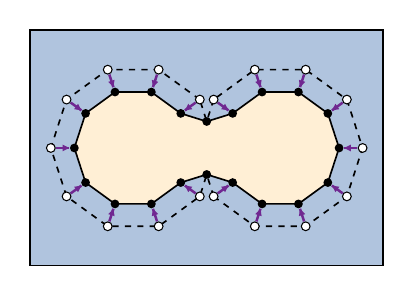}
    \\[0.6em]
    \includegraphics[width=0.42\textwidth]{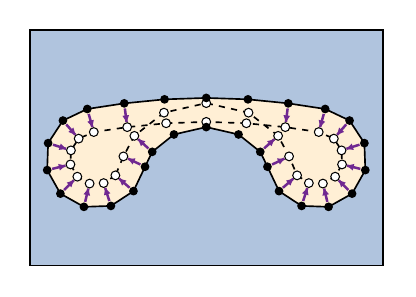} &
    \includegraphics[width=0.42\textwidth]{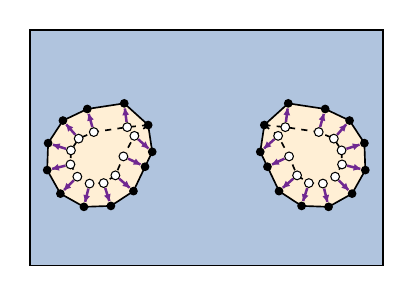}
    \end{tabular}
    \caption[Feature-size control through artificial topological changes]{Feature-size control by growth (top) and shrinkage (bottom). Left: markers move from initial positions (filled circles) to offset positions (open circles); arrows in the central overlap are omitted. Right: after topological reconstruction, surviving markers return to their initial positions, while new intersection markers remain fixed. Dashed contours show offset interfaces; solid contours show the initial (left) or final (right) interface. Both rows are schematic and use exaggerated displacements. The two operations are applied to both phases.}
    \label{fig:topo_control}
\end{figure}

The present framework therefore includes a feature-size control procedure based on artificial topological changes. The principle consists in temporarily displacing the markers by a distance $\delta$ in the interface normal direction before applying the topological reconstruction algorithm. There are two complementary operations. Growing a selected phase can close a thin intervening film; shrinking it can sever a thin neck. Each temporary offset is followed by topological reconstruction and restoration toward the original geometry. Existing markers are restored to their saved positions, while newly created markers remain fixed at their reconstructed positions. If an offset creates no contact, that pass leaves the interface unchanged. The implementation applies $+\delta$ and then $-\delta$ relative to the oriented marker normals; which sign grows a given phase depends on its left/right label. Features thinner than approximately $2\delta$ are targeted. Figure~\ref{fig:topo_control} illustrates the offset, reconstruction and return sequences for both growth and shrinkage, using exaggerated displacements to make the removal of small features visible.

This operation is not strictly volume preserving: removing features below the $2\delta$ scale induces a localized change in phase area. The parameter $\delta$ must therefore be regarded as both a geometrical cutoff and a modelling parameter.

This procedure provides a direct control over the minimal geometrical scales represented by the interface. In applications where thin films play an important physical role, $\delta$ can be chosen very small in order to delay topological transitions. Conversely, in problems dominated by larger flow structures, larger values of $\delta$ may be employed to simplify the interface geometry and improve the robustness of the mesh deformation procedure.

\section{X-Mesh: a mesh deformation algorithm}
\label{sec:xmesh}

The marker geometry is now known. We next fit the computational mesh to this interface and assign each element to its physical phase. The objective of the X-Mesh deformation procedure is to construct a computational mesh locally conforming to the marker-based interface representation while preserving the original mesh connectivity \cite{Moes2023XMeshStefan,Quiriny2024XMeshTwoPhase}. In contrast to classical interface-fitted approaches, the method does not rely on remeshing operations and instead accommodates arbitrarily distorted or degenerated elements near the interface.

In the other level-set-based X-Mesh formulation, the interface was represented directly on the computational mesh through the zero iso-contour of the level-set field. As a consequence, the smallest representable interface structures were intrinsically limited by the mesh resolution. In the present marker-based formulation, the interface is represented independently of the computational mesh and may therefore contain features significantly smaller than the local element size. This is particularly advantageous for the representation of thin filaments or films, where the X-Mesh framework naturally allows the introduction of nearly degenerated elements inside the structure without modifying the mesh topology.

The purpose of the deformation algorithm is therefore to interpolate the marker interface by selected mesh edges, so that the deformed mesh provides a body-fitted representation of the tracked geometry.

\subsection{Mesh deformation algorithm}

\begin{figure}[H]
    \centering
    \includegraphics[width=0.94\textwidth]{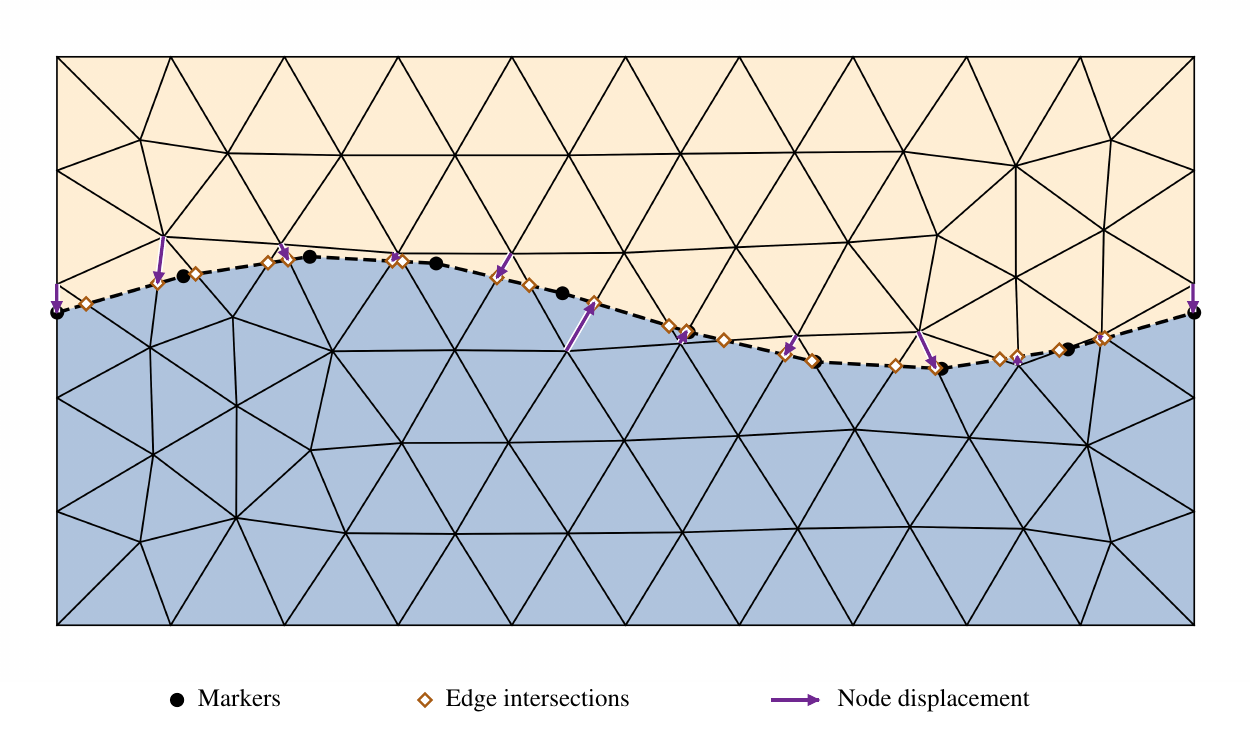}

    \vspace{0.5em}
    \includegraphics[width=0.94\textwidth]{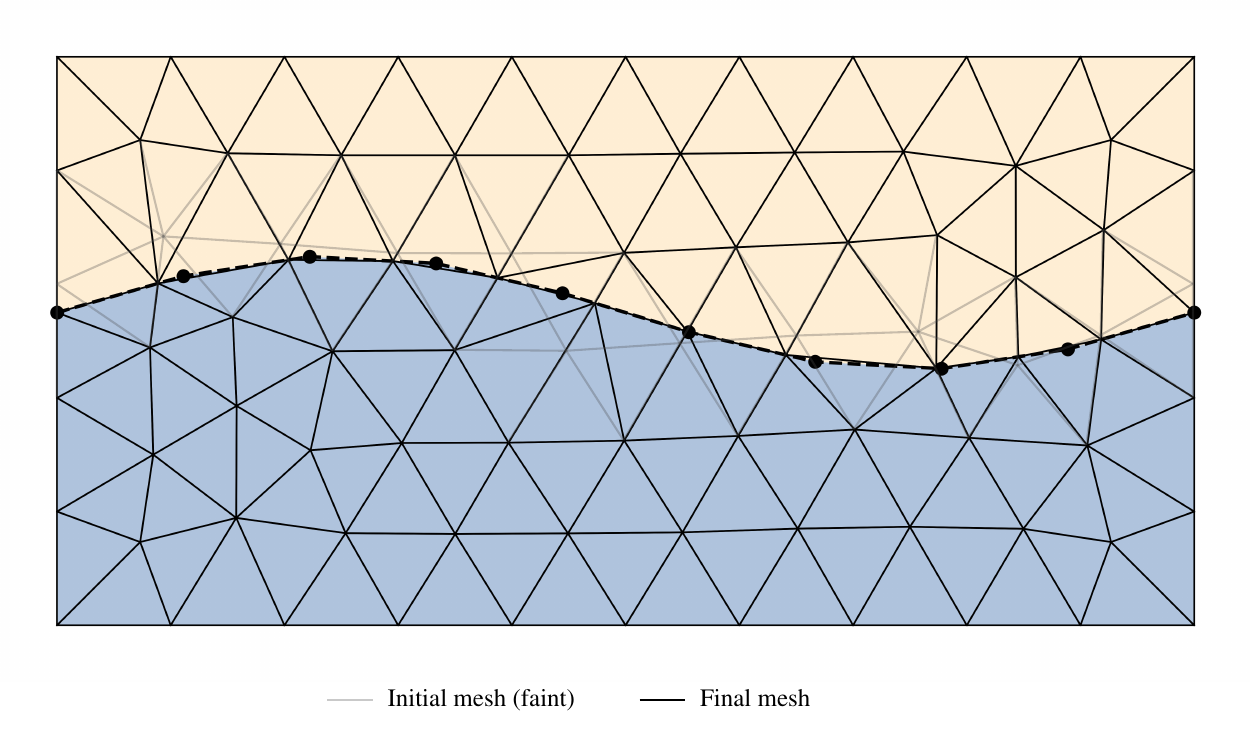}
    \caption[Mesh deformation for a sinusoidal interface]{Mesh deformation for a sinusoidal interface. Top: initial mesh, with edge--interface intersections marked by open diamonds and arrows showing the net node displacements towards the final configuration. Boundary contacts are shown by the endpoint markers. Bottom: final mesh, with the initial mesh superimposed in translucent gray; unchanged edges coincide. Filled black disks denote interface markers.}
    \label{fig:marker_defo}
    \label{fig:marker_defo_2}
    \label{fig:marker_defo_3}
    \label{fig:marker_defo_4}
\end{figure}

We start from an initial computational mesh together with a set of markers representing the interface. An example configuration is shown in Figure \ref{fig:marker_defo} (top), where the interface is represented by a sinusoidal marker chain embedded inside a non-conforming mesh.

Only admissible node displacements are authorized during the deformation process. In particular, corner nodes remain fixed, boundary nodes are constrained to move along the boundary, and node displacements leading to inverted elements are rejected.

The first stage of the algorithm consists in capturing zero-dimensional interface features. In practice, these features correspond to triple points or contact points between the interface and the domain boundary. The triangle containing the feature is identified and the closest admissible mesh node is displaced directly onto the marker position. The boundary arrows in Figure~\ref{fig:marker_defo} (top) show the displacement of mesh vertices onto the two contact markers.

The second stage consists in computing the intersections between mesh edges and marker segments. These intersections define admissible target locations toward which mesh nodes may be displaced. Using edge intersections restricts the candidate displacements; element inversion is checked separately for every move.

The open diamonds in Figure~\ref{fig:marker_defo} (top) mark edge--interface intersections on the initial mesh. Each node may possess several admissible targets, but only the closest one is retained. At this local interface-conforming stage, the candidate nodes are the endpoints of mesh edges intersected by a marker segment; all other nodes remain unchanged, apart from the nodes used to capture zero-dimensional features. Bulk nodes may subsequently move during the Winslow optimization described in Section~\ref{sec:winslow}.

Finally, within each pass, mesh vertices are sequentially displaced toward their corresponding targets. Nodes are processed according to their distance to the target position, starting from the closest ones. Furthermore, each mesh edge is used only once within a pass in order to avoid multiple nodes collapsing onto the same interface location.

The target list is computed at the start of a pass and is not recomputed after each individual node displacement. Each candidate move is nevertheless checked for element inversion on the mesh updated by the preceding moves. Once the pass is complete, edge--marker intersections are recomputed and a new pass begins. The iteration stops when no unresolved intersections remain or when no further admissible displacement can be made. The resulting mesh, illustrated in Figure \ref{fig:marker_defo_4} (bottom), becomes locally conforming to the marker interface while preserving the original mesh connectivity.

\Needspace{0.45\textheight}
\subsection{Degenerate elements and their treatment}
\label{sec:degenerate_elements}
\begin{figure}[H]
\centering
\includegraphics[width=0.78\textwidth]{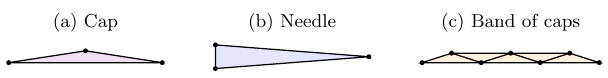}
\caption{Near-degenerate triangles and a band of caps; thicknesses are exaggerated.}
\label{fig:tfem_degeneracies}
\end{figure}

\begingroup
Since X-Mesh uses degenerate elements, we recall how their shape affects finite elements. In two dimensions, triangular elements degenerate into needles, with two angles close to $\pi/2$, or caps, with one angle close to $\pi$ (Figure~\ref{fig:tfem_degeneracies}). Bands of caps are the only such two-dimensional pattern that threatens FEM convergence: their coupled constraints can force a linear profile along the band and cause locking \cite{Quiriny2026TFEM}. The fitting algorithm is designed to minimize the formation of bands of caps and avoids them in almost all practical cases.
\endgroup

\begingroup
We nevertheless use the Tempered Finite Element Method (TFEM) \cite{Quiriny2026TFEM} as an additional safeguard against any remaining bands of caps. Replacing $1/J$ by $1/\max(J,J_{\min})$ in the stiffness contribution limits the excessive stiffness responsible for locking.
\endgroup

\begingroup
We set $J_{\min}=h^3/L$, where $h$ is the prescribed background mesh size and $L$ is the characteristic length of the solution.
\endgroup

\subsection{Phase identification on the computational mesh}
\label{sec:mesh_phase}

The algorithm maintains two coupled topological descriptions: the oriented marker partition, which defines the physical phases, and the set of interface-fitted mesh edges, which supports the finite-element discretization. %

Once the mesh has been deformed, each computational element must be assigned to one of the fluid phases to provide the phase representation required by the finite-element discretization. In the present work, the phase assignment is primarily performed through a breadth-first search (BFS) propagation algorithm.

We initialize the search from a triangle with an unambiguous phase. The winding number is evaluated at its three vertices and its barycenter: all four points must lie in the same phase, and none may lie on the interface. Starting from this seed, a FIFO queue propagates the phase to neighboring triangles across ordinary mesh edges. Interface edges act as barriers to copying the same phase; their oriented labels specify the phase on the opposite side. If a component remains unvisited, the same seed test is used to restart the search there.

In particular, the BFS strategy is more robust than a pure winding-number approach when the mesh deformation locally cuts regions of high curvature. Such situations may occur when elongated or highly distorted elements intersect the marker interface in a non-conforming manner. In these cases, the winding number evaluated at the barycenter of an element may incorrectly classify the element phase even though the interface represented on the mesh clearly indicates the opposite configuration.

The BFS strategy, however, requires the interface representation on the mesh to form a closed set of edges. If the mesh is too coarse compared to the smallest interface features, gaps may appear in the reconstructed interface. In that situation, the BFS propagation may leak through the gap and incorrectly color the entire domain with the same phase.

The discrete support is checked for gaps: a closed marker curve should have a closed support, while an open chain should have only its expected endpoints at walls or shared junctions. If a gap is detected, a local winding fallback classifies nearby elements from the winding number at their barycenters. Edges separating elements of different phases then become additional barriers for BFS propagation.

\subsection{Boundary management and triple points}
\label{sec:boundary_markers}

In the present framework, solid boundaries are represented using the same marker formalism as fluid interfaces. Boundary markers remain fixed on the domain boundary, except at triple points where the fluid interface meets the wall.

\begin{figure}[!ht]
    \centering
    \includegraphics[width=0.6\textwidth,trim={10cm 0cm 10cm 0cm},clip]{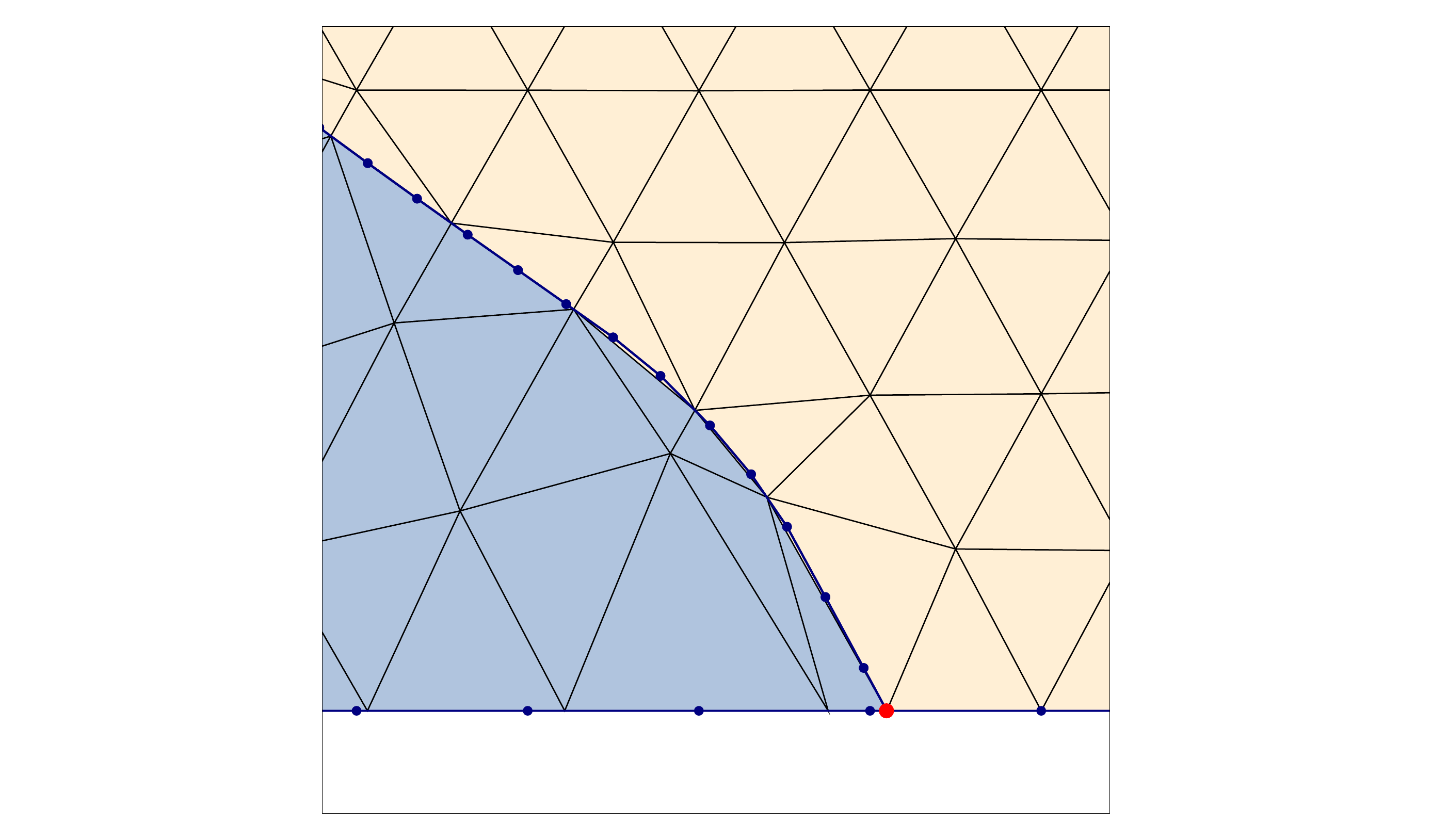}
    \caption{Close-up view of the mesh near a triple point. Blue disks denote interface and fixed wall markers, while the red disk identifies the moving fluid--fluid--solid contact point. The blue disks represent the same marker discretization as the black disks in Figure~\ref{fig:marker_defo_4}; their shape has no algorithmic significance.}
    \label{fig:triple_point_mesh}
\end{figure}

A boundary triple point is a shared endpoint of a fluid-interface chain and the adjacent wall segments. The feature-capture stage assigns a mesh vertex to this marker position (Figure~\ref{fig:triple_point_mesh}). This section describes its geometrical representation and ordering along the wall.

In the present work, wall impermeability is enforced while the tangential velocity remains free at the triple points. Their displacement is therefore obtained by projecting the local fluid velocity onto the boundary tangent direction. Triple points are then advected along the wall while the remaining boundary markers remain fixed. When reaching a corner of the domain, the projected velocity naturally drives the triple point along the adjacent boundary segment.

A contact point can slide past the position of a fixed marker along the wall. It remains on the wall throughout this motion; `crossing' here means overtaking a marker in the one-dimensional wall ordering. The boundary-marker list is therefore sorted by arclength after advection, and the wall segments adjacent to the contact point are updated to preserve consistent connectivity.

\subsection{Winslow energy minimization for bulk mesh deformation}
\label{sec:winslow}

In addition to the local interface-conforming deformation, nodes located in the bulk of the domain are also displaced in order to improve the overall mesh quality and increase the mesh resolution near the interface.

To achieve this objective, we employ the foldover-free mesh optimization strategy proposed in \cite{FoldoverFree}, based on the minimization of a regularized Winslow functional \cite{CharakhchyanIvanenko1997Winslow}. The implementation used in the present work relies on the corresponding optimization tools available in Gmsh \cite{geuzaine2009gmsh}.

The Winslow energy balances two competing objectives: preserving element areas and maintaining element angular quality. In the original formulation of \cite{FoldoverFree}, the target mesh corresponds to a tessellation of equilateral elements of uniform size. In the present work, this approach is extended by introducing a spatially varying target element size field.

More precisely, small target sizes are prescribed in the vicinity of the interface while larger target sizes are imposed farther away from it, together with a smooth transition region between the two scales. During the optimization process, boundary nodes remain fixed.

\begin{figure}[!ht]
    \centering
    \includegraphics[width=0.8\textwidth]{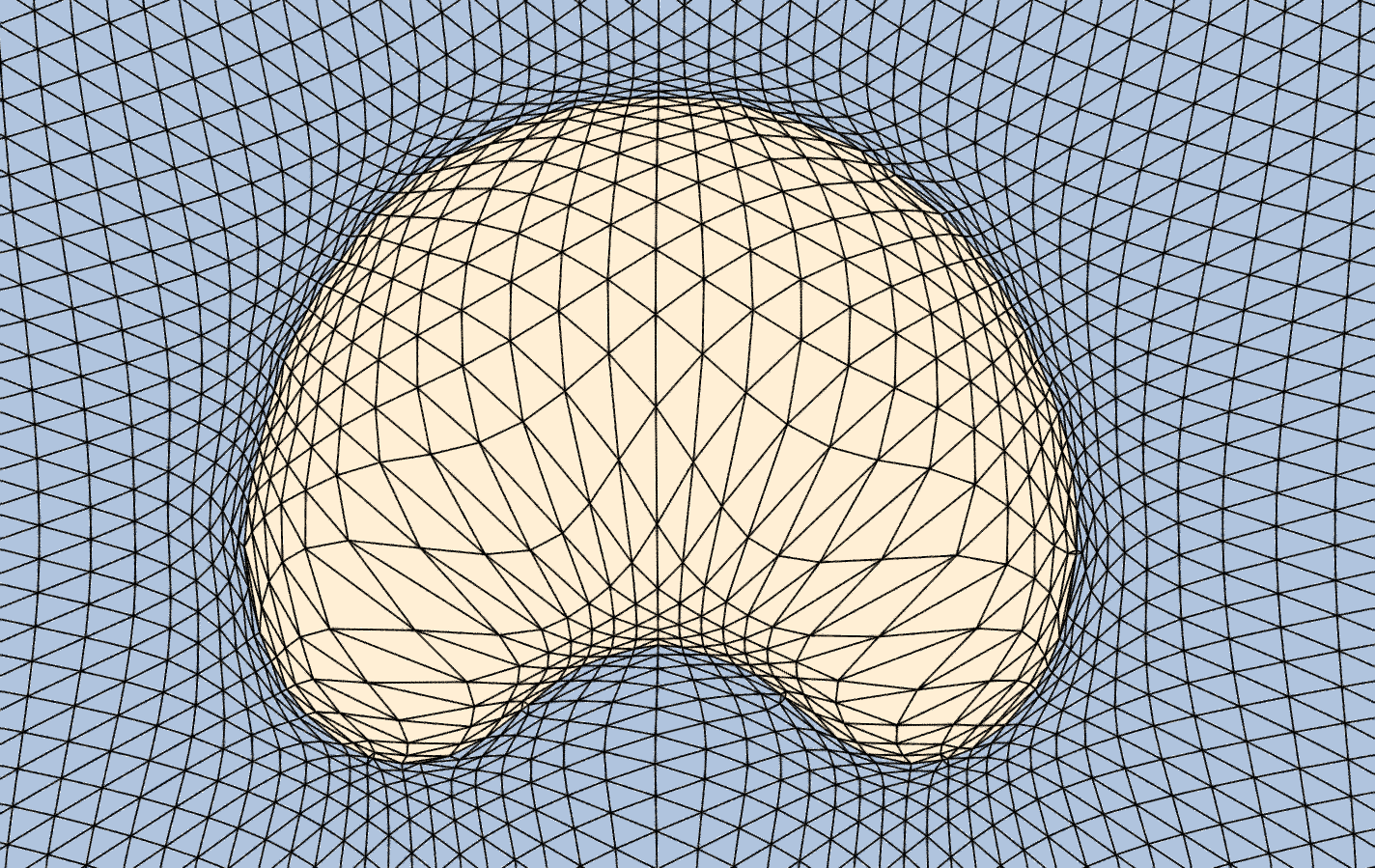}
    \caption{Example of mesh obtained after Winslow energy minimization in a bubble simulation (exaggerated deformation for visualization purposes).}
    \label{fig:winslow}
\end{figure}

An example of the resulting mesh is shown in Figure \ref{fig:winslow}. By redistributing mesh nodes toward the interface region, the Winslow optimization substantially improves the ability of the X-Mesh framework to represent thin films, boundary layers, and highly localized interfacial structures. 

\section{An ALE finite element solver with an interface-broken pressure}
\label{sec:fluid}

Let $\Omega=\Omega_0(t)\cup\Gamma(t)\cup\Omega_1(t)$ be occupied by two
immiscible incompressible Newtonian fluids, separated by the interface $\Gamma(t)=\partial\Omega_0(t)\cap\partial\Omega_1(t)$. For clarity, the equations below are written for two phases; the formulation applies to $N$ phases by summing over all phase domains and pairwise interfaces. The density $\rho_k$ and dynamic
viscosity $\mu_k$ are constant in each phase $\Omega_k(t)$. In an Arbitrary
Lagrangian--Eulerian (ALE) description, the governing equations read
\begin{align}
\rho_k\left(
\left.\partial_t\mathbf{u}\right|_{\boldsymbol\chi}
+(\mathbf{u}-\mathbf{w})\cdot\nabla\mathbf{u}
\right)
-\nabla\cdot\left(2\mu_k\mathbf{D}(\mathbf{u})\right)
+\nabla p_k
&=\rho_k\mathbf{g}
&&\text{in }\Omega_k(t),
\label{eq:ale_momentum}
\\
\nabla\cdot\mathbf{u}&=0
&&\text{in }\Omega_k(t),
\label{eq:ale_incompressibility}
\end{align}
Here $\boldsymbol\chi$ denotes the fixed coordinates of the reference ALE mesh: $\left.\partial_t\right|_{\boldsymbol\chi}$ follows that moving mesh. The mesh velocity is $\mathbf{w}$, and
$\mathbf{D}(\mathbf{u})=(\nabla\mathbf{u}+\nabla\mathbf{u}^{T})/2$. The
velocity is continuous across the interface, whereas the pressure and viscous
stress may be discontinuous. For example, $[p]=p_1|_\Gamma-p_0|_\Gamma$ is the difference of the two pressure traces. The same convention applies componentwise to the stress tensor. We take $\mathbf n$ from phase 0 to phase 1, and define the signed curvature consistently with the following stress balance:
\begin{equation}
\left[
-p\mathbf{I}+2\mu\mathbf{D}(\mathbf{u})
\right]\mathbf{n}
=\sigma\kappa\mathbf{n}\qquad\text{on }\Gamma(t).
\label{eq:jump_condition}
\end{equation}
In the present formulation, condition~\eqref{eq:jump_condition} is not
enforced through a prescribed curvature-dependent pressure correction. The
pressure jump is instead an unknown of the fluid problem, and surface-tension forces are obtained from the first variation of an interfacial membrane energy \cite{Leblond2013Membrane}.

The X-Mesh deformation keeps a fixed mesh connectivity while making a set of
mesh edges coincide with $\Gamma(t)$. Equations~\eqref{eq:ale_momentum} and
\eqref{eq:ale_incompressibility} are therefore solved on the moving,
interface-fitted triangulation $\mathcal{T}_h(t)$ within the ALE framework
\cite{Hirt1974ALE,Donea1982ALEFE}. Extremely small element Jacobians generated
by the deformation are regularized with TFEM
described in Section~\ref{sec:degenerate_elements}.

\subsection{Pressure space continuous within each phase}
\label{sec:broken_pressure}

Both velocity and pressure use piecewise linear approximations. The velocity
space remains globally continuous,
\begin{equation}
\mathbf{V}_h=\left[P_1^c(\mathcal{T}_h)\right]^2,
\end{equation}
whereas the pressure is continuous only within each phase:
\begin{equation}
Q_h^\Gamma=
\left\{
q_h\,;\;
q_h|_{\Omega_k}\in P_1^c(\mathcal{T}_h|_{\Omega_k}),\quad k=0,1
\right\}.
\label{eq:broken_pressure_space}
\end{equation}
Thus, a vertex whose support contains elements from both phases carries two
pressure degrees of freedom, one for each trace of $p_h$ on $\Gamma$. All
other vertices retain a single pressure degree of freedom. If $N_\Gamma$
denotes the number of interface vertices, the pressure space contains
$N_p=N_v+N_\Gamma$ degrees of freedom, while the velocity numbering is
unchanged. This enrichment within each phase provides a sharp representation of the
physical pressure discontinuity without a Lagrange multiplier or an imposed
jump value.

For more than two phases, let $m_a$ be the number of distinct fluid phases represented by the elements incident to vertex $a$. Assigning one pressure trace to each incident phase gives $N_p=\sum_{a=1}^{N_v}m_a$. The preceding expression $N_p=N_v+N_\Gamma$ is its two-phase specialization. An interior triple junction has $m_a=3$ and therefore carries three pressure degrees of freedom.

Because the equal-order pair $\mathbf{V}_h\times Q_h^\Gamma$ does not satisfy
the discrete inf--sup condition, pressure-stabilizing Petrov--Galerkin (PSPG)
stabilization is applied elementwise \cite{HughesFrancaBalestra1986PSPG}.
Streamline Upwind/Petrov--Galerkin (SUPG) stabilization is also included for
the convection-dominated regimes \cite{BrooksHughes1982SUPG}. A single
volume-weighted mean-pressure constraint is imposed over $Q_h^\Gamma$; the two
phases are not constrained independently, so their relative pressure level
remains part of the solution.

\subsection{Discrete ALE formulation}

Let $\mathbf{x}_h^n$ be the nodal coordinates of the mesh fitted to
$\Gamma_h^n$. At the beginning of the time step, the marker interface is
advected explicitly with $\mathbf{u}_h^n$, and the mesh is subsequently
deformed to fit the resulting interface $\Gamma_h^{n+1}$. The ALE mesh
velocity used in the fluid solve is therefore the velocity of this current
deformation,
\begin{equation}
\mathbf{w}_h^{n+1}=
\frac{\mathbf{x}_h^{n+1}-\mathbf{x}_h^n}{\Delta t}.
\label{eq:mesh_velocity}
\end{equation}
Its nodal values are extended piecewise linearly over the updated mesh
$\mathcal{T}_h^{n+1}$. Because the connectivity is fixed, the old fluid velocity
degrees of freedom can be carried with their mesh nodes. We denote by
$\widetilde{\mathbf{u}}_h^n$ the corresponding ALE transport of the old
velocity to $\mathcal{T}_h^{n+1}$, defined nodally by
\begin{equation}
\widetilde{\mathbf{u}}_h^n(\mathbf{x}_{h,a}^{n+1})
=\mathbf{u}_h^n(\mathbf{x}_{h,a}^n),
\qquad a=1,\ldots,N_v.
\label{eq:ale_velocity_transport}
\end{equation}

After the mesh deformation, the elements are classified using the updated
marker geometry as described in Section~\ref{sec:mesh_phase}. Thus the phase
domains $\Omega_{k,h}^{n+1}$, the material coefficients, the fitted interface
$\Gamma_h^{n+1}$, and the broken pressure space
$Q_h^{\Gamma^{n+1}}$ used in the fluid problem all correspond to time
$t^{n+1}$; none of these quantities is lagged from the previous element
partition. In the following, $\mathbf{V}_h^{n+1}$ denotes the continuous
velocity space on $\mathcal{T}_h^{n+1}$, and
$\mathbf{V}_{h,0}^{n+1}$ its subspace satisfying the homogeneous essential
boundary conditions. An implicit Euler discretization is used for the fluid
variables, and the convective velocity is the current ALE relative velocity
$\mathbf{u}_h^{n+1}-\mathbf{w}_h^{n+1}$.

\begingroup
For readability in the following equations, the discrete fields, test functions, domains and interface are all evaluated on $\mathcal T_h^{n+1}$; the indices $h$ and $n+1$ are suppressed. We write $\widetilde{\mathbf u}^{n}$ for the transported old velocity. The pressure contribution has two parts: its ordinary gradient inside each phase and the jump of its traces on the interface. Their combined action is
\begin{equation}
b(\mathbf v,p)=
\sum_{k=0}^1\int_{\Omega_k}\nabla p_k\cdot\mathbf v\,\mathrm dx
+\int_\Gamma[p]\,\mathbf v\cdot\mathbf n\,\mathrm ds.
\label{eq:broken_pressure_momentum}
\end{equation}
The incompressibility contribution is the negative adjoint,
\begin{equation}
-b(\mathbf u,q)=-\sum_{k=0}^1\int_{\Omega_k}\mathbf u\cdot\nabla q_k\,\mathrm dx
+\int_\Gamma(q_0-q_1)\,\mathbf u\cdot\mathbf n\,\mathrm ds,
\label{eq:broken_pressure_incompressibility}
\end{equation}
up to external-boundary contributions. The traces $q_0$ and $q_1$ are independent, just as the pressure traces are. Their interface coupling and the single global mean-pressure constraint are retained. The derivative of $b(\mathbf v,p)$ with respect to both pressure traces is assembled in the Newton matrix.

The symbols $\delta E_\Gamma(\mathbf v)$ and $\delta E_w(\mathbf v)$ denote the first variations of the interfacial and wetted-wall energies in the direction $\mathbf v$. Their explicit segment formulas are given in Section~\ref{sec:surface_tension}. With this notation, the complete residual is
\begin{multline}
\sum_{k=0}^1\int_{\Omega_k}
\rho_k\left(\frac{\mathbf u-\widetilde{\mathbf u}^{\,n}}{\Delta t}
+((\mathbf u-\mathbf w)\cdot\nabla)\mathbf u-\mathbf g\right)\cdot\mathbf v
+2\mu_k\mathbf D(\mathbf u):\mathbf D(\mathbf v)\,\mathrm dx
\\
+b(\mathbf v,p)-b(\mathbf u,q)+\mathcal S(\mathbf v,q)
+\delta E_\Gamma(\mathbf v)+\delta E_w(\mathbf v)=0.
\label{eq:weak_ns_broken}
\end{multline}
We seek $(\mathbf u,p)\in\mathbf V_h^{n+1}\times Q_h^{\Gamma^{n+1}}$ and test with $(\mathbf v,q)\in\mathbf V_{h,0}^{n+1}\times Q_h^{\Gamma^{n+1}}$. The stabilization keeps the same elementwise momentum residual $\mathcal R_{m,K}$:
\begin{equation}
\mathcal S(\mathbf v,q)=\sum_{K\in\mathcal T_h^{n+1}}\int_K
\tau_K\big((\mathbf u-\mathbf w)\cdot\nabla\mathbf v\big)\cdot\mathcal R_{m,K}
+\frac{\tau_K}{\rho_K}\nabla q\cdot\mathcal R_{m,K}\,\mathrm dx,
\label{eq:supg_pspg}
\end{equation}
\begin{equation}
\tau_K=\left[\left(\frac{2}{\Delta t}\right)^2
+\left(\frac{2\|\mathbf u-\mathbf w\|}{h_K}\right)^2
+\left(\frac{4\nu_K}{h_K^2}\right)^2\right]^{-1/2},
\qquad\nu_K=\frac{\mu_K}{\rho_K}.
\label{eq:stabilization_parameter}
\end{equation}
The element size $h_K$ is retained explicitly.
\endgroup
All occurrences of $\mathbf{u}_h^{n+1}$, including those in the convective
term and in the stabilization residual, are included in the nonlinear
problem. The resulting fully implicit Navier--Stokes system is solved by
Newton--Raphson iterations on the already updated geometry.

\subsection{Variational membrane model for surface tension}
\label{sec:surface_tension}

Following Leblond et al.~\cite{Leblond2013Membrane}, the fitted interface is
treated as a massless membrane carrying an isotropic tension $\sigma$. For the
constant surface tension considered here, its two-dimensional energy is
\begin{equation}
E_\Gamma(\mathbf{x})=
\sigma\int_\Gamma\mathrm{d}s.
\label{eq:membrane_energy}
\end{equation}
Let $\Gamma_h$ consist of interface edges $e=(a,b)$ of length $\ell_e$, with
unit tangent
\begin{equation}
\mathbf{t}_e=\frac{\mathbf{x}_b-\mathbf{x}_a}{\ell_e}.
\end{equation}
The discrete energy and its first variation are
\begin{align}
E_{\Gamma,h}(\mathbf{x}_h)
&=\sigma\sum_{e\subset\Gamma_h}\ell_e,
\\
\delta E_{\Gamma,h}(\mathbf{v}_h)
&=\sigma\sum_{e=(a,b)\subset\Gamma_h}
\mathbf{t}_e\cdot(\mathbf{v}_b-\mathbf{v}_a).
\label{eq:membrane_variation}
\end{align}
Thus, one edge contributes the nodal residuals
$-\sigma\mathbf{t}_e$ and $+\sigma\mathbf{t}_e$ at $a$ and $b$,
respectively. Tangential contributions cancel at an interior vertex of a
locally straight interface, whereas their resultant on a curved polygonal
interface provides the surface-tension force. No discrete curvature field is required, and the pressure jump remains an unknown of the coupled problem. The jump $[p_h]$
emerges from the balance between~\eqref{eq:broken_pressure_momentum} and
\eqref{eq:membrane_variation}; in particular, the Laplace pressure is recovered
at static equilibrium.

\begingroup
The same energetic formulation applies at a solid wall. Let $S$, $L$ and $G$ denote the solid, the selected wetting fluid and the other fluid, respectively, with interfacial tensions $\gamma_{SL}$, $\gamma_{SG}$ and $\gamma_{LG}=\sigma$. For a fixed total wall length, the wall energy is
\begin{equation}
E_w=\gamma_{SL}L_{SL}+\gamma_{SG}L_{SG}
=(\gamma_{SL}-\gamma_{SG})L_w+C_w,
\label{eq:young_wall_energy}
\end{equation}
where $L_w=L_{SL}$ and $C_w=\gamma_{SG}(L_{SL}+L_{SG})$ is constant. Its discrete first variation is
\begin{equation}
\delta E_{w,h}(\mathbf v_h)=(\gamma_{SL}-\gamma_{SG})
\sum_{e=(a,b)\subset\Gamma_{w,h}}
\mathbf t_e\cdot(\mathbf v_b-\mathbf v_a),
\label{eq:jon_wall_variation}
\end{equation}
where $\Gamma_{w,h}$ is the wetted wall and admissible displacements are tangential to the wall. Contributions cancel at interior wall vertices, leaving the contact-point forces. Combined with the fluid-interface tension, their tangential equilibrium gives Young's relation $\gamma_{SG}-\gamma_{SL}=\gamma_{LG}\cos\theta_Y$, where $\theta_Y$ is measured inside $L$. Choosing these tensions is therefore equivalent to prescribing that equilibrium angle, without constraining it kinematically. The contact points move with the free tangential velocity along the impermeable wall. The present model uses constant interfacial tensions and includes neither Marangoni stresses nor contact-line friction.
\endgroup

For $N$ fluids, the membrane energy is summed over pairwise interfaces: $E_\Gamma=\sum_{i<j}\sigma_{ij}\int_{\Gamma_{ij}}\mathrm{d}s$. At a free triple junction, the three incident marker chains share a common node and the membrane endpoint forces add at the same fitted mesh vertex. Their equilibrium is the Neumann balance stated in Eq.~\eqref{eq:lens_neumann}. This construction is exercised by the interior triple junctions in Section~\ref{sec:liquid_lenses}.

The membrane geometry used during the step is
$\Gamma_h^{n+1}$, obtained from the explicit marker advection and the
subsequent mesh deformation. This geometry is known before the Navier--Stokes
solve and is kept fixed during the Newton iterations; consequently, its
geometric tangent is not included in the Newton matrix. The capillary force is
therefore evaluated on the updated interface but remains explicit with respect
to the interface position. The computations use the capillary-wave restriction
\cite{popinet_fronttracking}
\begin{equation}
\Delta t\leq C_\sigma
\sqrt{\frac{(\rho_0+\rho_1)h_\Gamma^3}{4\pi\sigma}},
\qquad C_\sigma<1,
\label{eq:capillary_time_step}
\end{equation}
where $h_\Gamma$ is the smallest current interface-edge length.

\subsection{Coupling strategy}
\label{sec:coupling_strategy}

The fluid, marker, and mesh problems are coupled by an
explicit-geometry/implicit-fluid partitioned algorithm. Contrary to a
fluid-first update, the interface and mesh are advanced before the
Navier--Stokes problem is solved. Starting from the state at $t^n$, the marker
positions are first advanced explicitly with the known velocity
$\mathbf{u}_h^n$, using the second-order Runge--Kutta scheme of
Section~\ref{sec:interface}. Marker redistribution, topology control, and the
management of boundary markers and triple points are then applied to this
predicted interface.

The fixed-connectivity mesh is next deformed so that selected edges conform to
the updated marker interface. The phase of every element is identified only
after this deformation. Hence the phase partition entering the fluid residual
is $\{\Omega_{0,h}^{n+1},\Omega_{1,h}^{n+1}\}$, and the density, viscosity,
interface edges, and broken pressure degrees of freedom are all those of the
updated configuration. The mesh velocity is then computed from the same
$n\to n+1$ deformation using~\eqref{eq:mesh_velocity}. Finally,
$(\mathbf{u}_h^{n+1},p_h^{n+1})$ is obtained from the fully implicit nonlinear
Navier--Stokes problem~\eqref{eq:weak_ns_broken}. The interface geometry is
fixed during this last solve, but the velocity, pressure, convection, and
stabilization residual are treated implicitly.\\
\\
\begin{algorithm}[H]
\DontPrintSemicolon
\KwIn{Marker interface $\Gamma_M^n$, fitted mesh coordinates
$\mathbf{x}_h^n$, and velocity $\mathbf{u}_h^n$}
\KwOut{$\Gamma_M^{n+1}$, $\mathbf{x}_h^{n+1}$,
$\mathbf{u}_h^{n+1}$, and $p_h^{n+1}$}
\vspace{0.2cm}
Advect $\Gamma_M^n$ explicitly with $\mathbf{u}_h^n$ using RK2 to obtain
$\Gamma_M^{n+1}$\;
Apply marker redistribution, topology control, and boundary/triple-point
updates\;
Deform the fixed-connectivity mesh with X-Mesh:
$\mathbf{x}_h^n\mapsto\mathbf{x}_h^{n+1}$, fitted to
$\Gamma_M^{n+1}$\;
Identify the phase of each $K\in\mathcal{T}_h^{n+1}$ and construct
$\Omega_{k,h}^{n+1}$ and $\Gamma_h^{n+1}$\;
Build the updated broken pressure space $Q_h^{\Gamma^{n+1}}$\;
Compute
$\mathbf{w}_h^{n+1}=(\mathbf{x}_h^{n+1}-\mathbf{x}_h^n)/\Delta t$\;
Solve the fully implicit Navier--Stokes system for
$(\mathbf{u}_h^{n+1},p_h^{n+1})$ by Newton--Raphson iterations\;
\vspace{0.2cm}
\caption{One time step of the partitioned marker--mesh--fluid coupling.}
\label{alg:marker_mesh_fluid_coupling}
\end{algorithm}\vspace{0.5cm}
If the element partition changes after an X-Mesh or topology update, the
broken-pressure space is rebuilt while the velocity degrees of freedom are
left unchanged. Since pressure is an instantaneous Lagrange multiplier rather
than an evolved state variable, no pressure field is transferred between the
two discrete spaces. All pressure degrees of freedom are initialized to zero
before the Newton solve on the updated mesh.

Zero pressure is only the initial guess; pressure is solved together with velocity. Since each Newton system is solved directly here, pressure transfer is not needed to initialize an iterative linear solver.

\newpage
\section{Numerical results}
\label{sec:results}

This section assesses the marker-based X-Mesh formulation through a sequence of complementary test cases. The vortex-in-a-box benchmark first isolates the geometrical part of the method, namely marker advection and mesh conformity under severe reversible stretching. The Rayleigh--Taylor instability then activates the full two-phase coupling in a configuration where the interface develops long filaments and topology changes, while a three-phase variant assesses the oriented phase-partition representation during strongly mixed motion. The two rising-bubble benchmarks of Hysing et al.~\cite{hysing2009bubble} provide quantitative comparisons with standard reference data in moderately and strongly deformed regimes. Finally, static sessile drops and capillary rise assess the membrane formulation at moving fluid--fluid--solid triple points. Three-fluid lenses then test the balance of unequal interfacial tensions at free triple junctions, with junction angles measured from circular fits of the marker interfaces.

\subsection{Vortex in a box}
\label{sec:vortex}

\begin{figure}[H]
\centering
\includegraphics[width=0.92\textwidth]{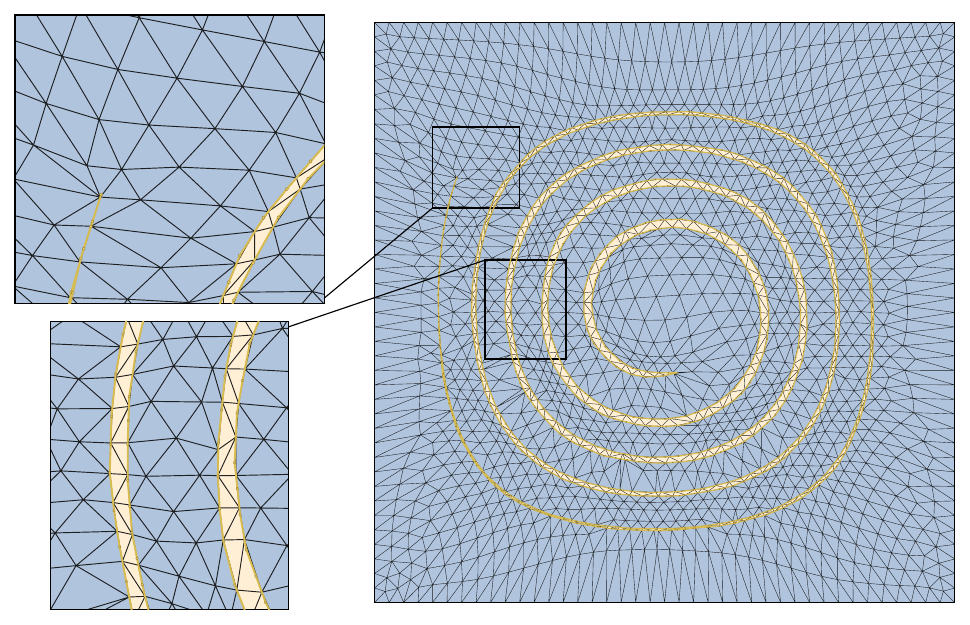}
\caption{Vortex-in-a-box advection test. The global view is complemented by two magnifications showing the marker-tracked interface and the locally conforming X-Mesh deformation near thin stretched structures.}
\label{fig:vortex_box}
\end{figure}

The first test consists in advecting a closed interface in a vortical velocity field. This benchmark is commonly used to assess the ability of interface-capturing methods to preserve thin structures under large deformations \cite{LeVeque1996VortexAdvection}. Here it is used as a geometrical stress test for the X-Mesh deformation algorithm with the marker representation. The domain is the unit square $\Omega=(0,1)^2$, and the initial interface is a circle of radius $0.15$ centered at $(0.5,0.75)$. The velocity field is given by
\begin{equation}
\label{eq:vortex_velocity}
\mathbf{u}(x,y) = \begin{pmatrix}
\sin(2\pi y)\sin^2(\pi x) \\
-\sin(2\pi x)\sin^2(\pi y)
\end{pmatrix}.
\end{equation}
Figure~\ref{fig:vortex_box} shows the deformed mesh together with two local magnifications. Although some interfacial features are smaller than the characteristic length of the background mesh, the conforming mesh captures the local geometry through highly anisotropic and locally degenerated elements. The zooms highlight two important properties of the method: thin ligaments can be represented below the characteristic mesh size, and the mesh remains conforming without remeshing or changes in connectivity.

The pointed end in the upper-left magnification is represented by very thin elements generated by mesh deformation. Their small thickness does not indicate local mesh refinement. Since the velocity is prescribed in this test, the figure demonstrates geometric resolution of the tip, rather than accuracy of a fluid solution at that scale.

The result confirms the expected role of the explicit marker discretization: the geometrical resolution of the front is not limited to one marker per background element. Instead, the mesh deformation introduces the degeneracies required to conform to the transported front. This behavior is central for $N$-phase simulations in which thin liquid films must be retained until an explicit feature-size or topology criterion is activated.

\subsection{Rayleigh-Taylor instability}
\label{sec:rayleigh_taylor}

The method is further assessed on the classical two-dimensional Rayleigh-Taylor instability benchmark. The computational domain is a vertical channel of width $W$ and height $4W$. Free slip boundary conditions are imposed on the four walls of the channel. The heavy fluid is initially placed above the light fluid and the interface is perturbed by a single sinusoidal mode of amplitude $0.1W$. The physical parameters are chosen to match the reference configuration of He et al.~\cite{HeChenZhang1999RTI}, with Atwood number $A_t=0.5$ and Reynolds number $Re=\sqrt{Wg}W/\nu=256$. Time is made dimensionless using the characteristic time scale $\sqrt{W/g}$.

\begin{figure}[H]
\centering
\includegraphics[width=\textwidth]{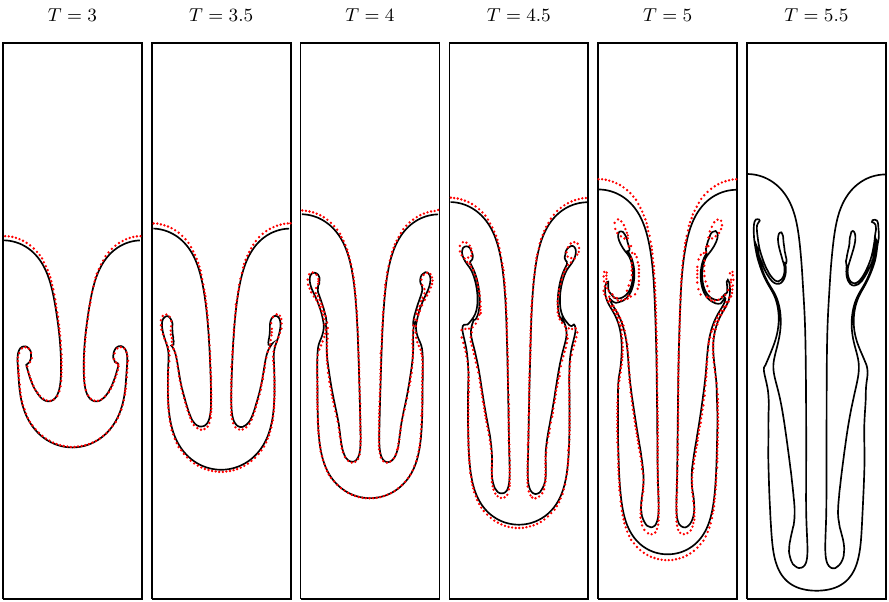}
\caption{Rayleigh--Taylor instability for $h=0.0125$ at dimensionless times $T=3$, $3.5$, $4$, $4.5$, $5$, and $5.5$. Red points show the available reference interfaces without connecting the raw coordinates. They are available here only up to $T=5$; no reference points are plotted at $T=5.5$.}
\label{fig:rt_finest_time_sequence}
\end{figure}

Figure~\ref{fig:rt_finest_time_sequence} shows the computed interfaces at dimensionless times $T=3,3.5,4,4.5,5$, and $5.5$, with the reference interfaces of He et al. overlaid for the first five times. No digitized reference coordinates at $T=5.5$ are included in the comparison dataset; the last panel therefore shows only the present computation. Overall, the simulation reproduces the main features of the Rayleigh-Taylor evolution: the downward motion of the heavy-fluid spike, the upward displacement of the light-fluid bubble, and the progressive roll-up of the lateral structures. The agreement is not exact, in particular at later times where the position and curvature of the secondary rolled-up structures differ from the reference data. These discrepancies are expected, since this regime is sensitive to the numerical treatment of the interface, to resolution, and to the amount of numerical diffusion or regularization. Nevertheless, the comparison remains qualitatively satisfactory: the global morphology, the symmetry of the instability, and the main penetration dynamics are captured without remeshing, while the thin interfacial structures remain explicitly represented by the marker-based conforming mesh.

\begin{figure}[H]
\centering
\includegraphics[width=\textwidth]{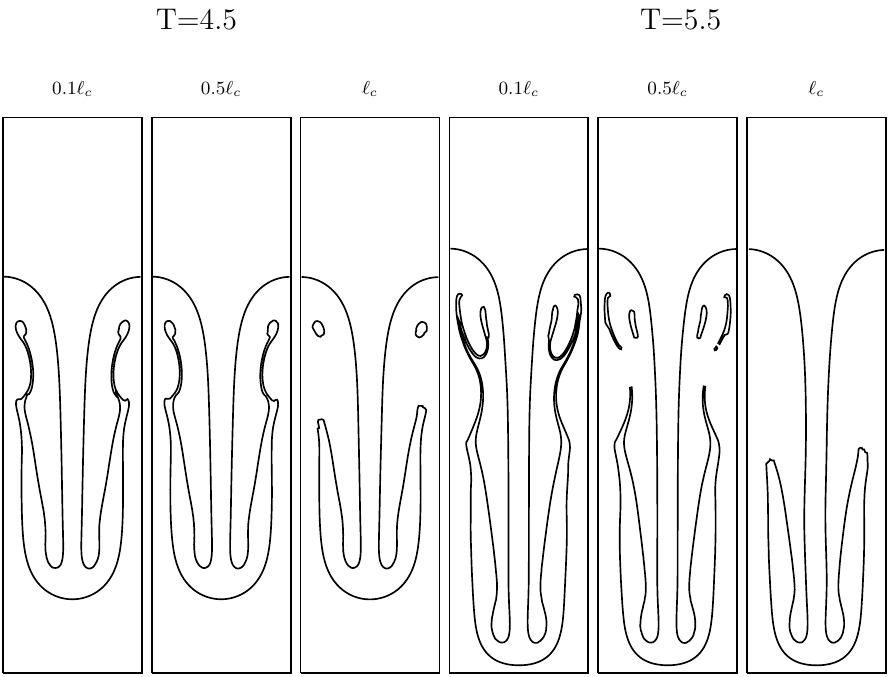}
\caption{Rayleigh--Taylor instability for $h=0.0125$ with offset distances $\delta=0.1\ell_c$, $0.5\ell_c$, and $\ell_c$, where $\ell_c=h$ and the corresponding feature-thickness cutoff is approximately $2\delta$, at dimensionless times $T=4.5$ and $T=5.5$.}
\label{fig:rt_cut_comparison}
\end{figure}

Figure~\ref{fig:rt_cut_comparison} illustrates the role of the offset distance $\delta$ used for feature-size control at times $T=4.5$ and $5.5$. In the numerical figures, $\ell_c$ denotes the prescribed background mesh size $h$, and the labels $0.1\ell_c$, $0.5\ell_c$, and $\ell_c$ specify $\delta$. The associated removed feature thickness is approximately $2\delta$, as described in Section~\ref{sec:interface}. Reducing the cut length from $\ell_c$ to $0.1\ell_c$ delays artificial topological changes and preserves thinner interfacial structures for a longer time. Conversely, larger cut lengths remove small ligaments earlier and lead to smoother rolled-up regions. This behavior is consistent with the intended role of the topology-control length: it is an explicit modelling parameter, not an accidental consequence of mesh resolution or numerical diffusion.

\begin{figure}[H]
\centering
\includegraphics[width=0.6\textwidth]{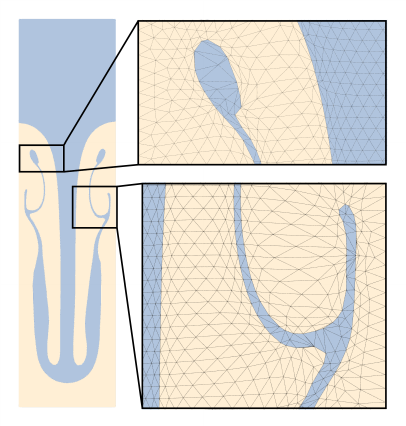}
\caption{Zoom on the Rayleigh-Taylor instability. The mesh deformation allows the finite element discretization to represent the thin interfacial structures.}
\label{fig:zoomRT}
\end{figure}

The local view in Figure~\ref{fig:zoomRT} highlights the mesh-deformation mechanism that makes this representation possible. The X-Mesh deformation aligns finite element edges with the transported interface, so that thin structures are represented sharply without spreading the material jump over a finite-thickness transition layer: each element belongs to one phase, and the density and viscosity change discontinuously across the fitted interface. The zooms also illustrate the effect of the Winslow-based optimization introduced above: mesh nodes are redistributed toward the interface, which improves the local resolution of curved and elongated features while preserving the mesh connectivity. As a result, interfacial structures smaller than the characteristic size of the background mesh can still be captured by locally anisotropic and nearly degenerated elements. Such features would typically be lost or strongly regularized in a level-set representation defined on the same background mesh resolution.

\subsection{Three-phase Rayleigh--Taylor instability}
\label{sec:rayleigh_taylor_three_phase}

The oriented phase-partition representation is next assessed with the three-phase Rayleigh--Taylor benchmark reported by Garoosi et al.~\cite{Garoosi2022VOF} and subsequently considered with an interface-fitted PFEM discretization by Ruyffelaere et al.~\cite{Ruyffelaere2026TopologyPFEM}. The domain is $\Omega=[0,H]\times[0,3H]$, with $H=1$, and initially contains three horizontal layers of increasing density, $\rho_1=1$, $\rho_2=2$, and $\rho_3=4$ from bottom to top. Both interfaces are perturbed according to
\begin{equation}
\Gamma_{1/2}(x): y=H+0.1H\cos(2\pi x/H),
\qquad
\Gamma_{2/3}(x): y=2H+0.1H\cos(2\pi x/H).
\end{equation}
The kinematic viscosity is uniform, $\nu=0.01$, so the dynamic viscosities in the three phases are $\mu_i=(0.01,0.02,0.04)$. Gravity is $\bm{g}=(0,-17.64)$ and both surface-tension coefficients vanish. This benchmark therefore isolates inertial, viscous, and topological multiphase effects without a capillary contribution. The computation uses a background size $h=1/83$, a topology-control length $\delta=1/830=0.1h$, and a time step $\Delta t=1.25\times10^{-3}$.

\begin{figure}[H]
\centering
\includegraphics[width=\textwidth]{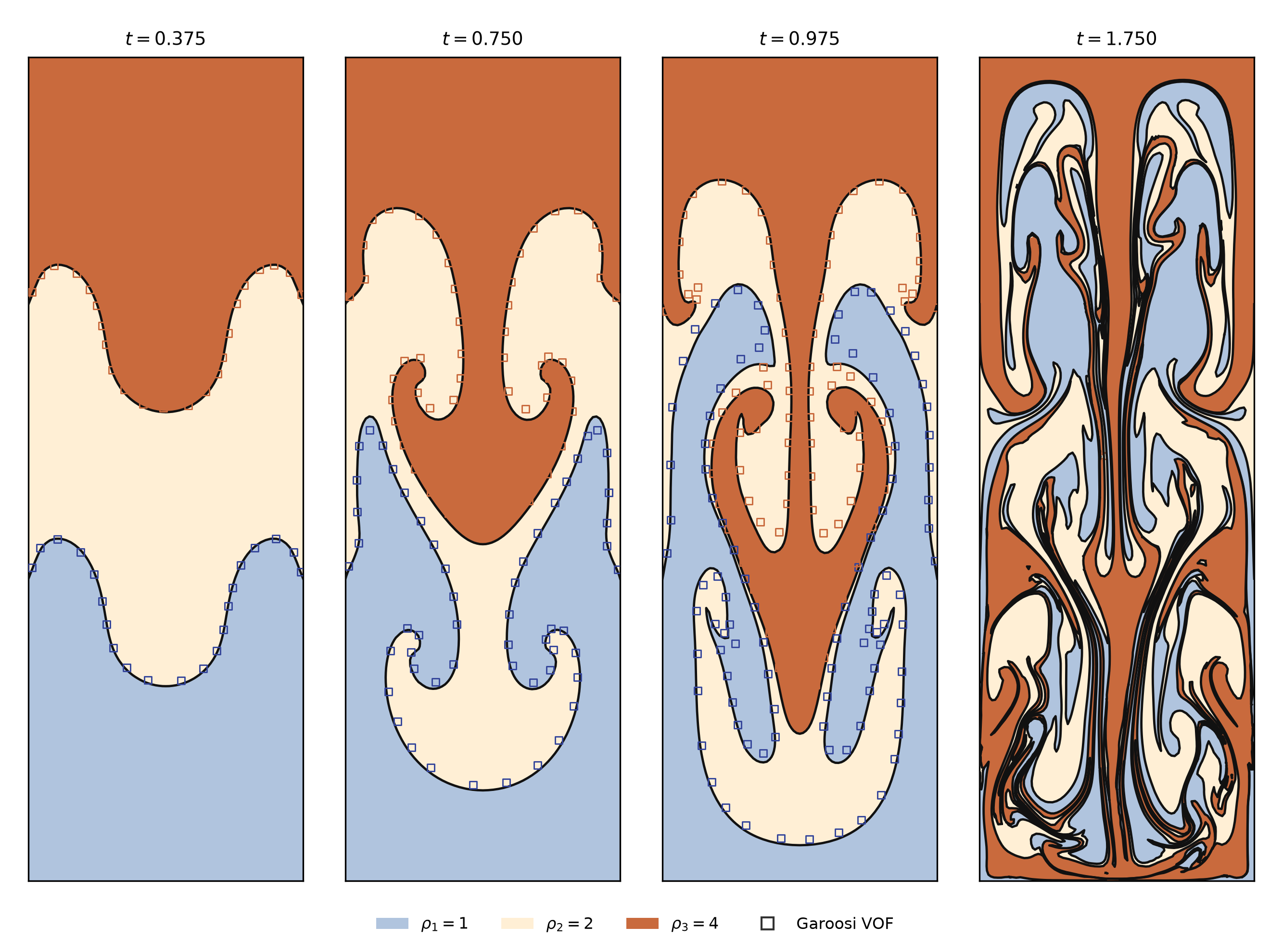}
\caption{Three-phase Rayleigh--Taylor instability at physical times $t=0.375$, $0.750$, $0.975$, and $1.750$. The phase densities are indicated by the filled colours and the mesh is omitted for clarity. In the first three panels, hollow squares show the lower and upper VOF interfaces reported by Garoosi et al.~\cite{Garoosi2022VOF}.}
\label{fig:rt_three_phase_shapes}
\end{figure}

Figure~\ref{fig:rt_three_phase_shapes} compares the interfaces obtained with the present approach with the VOF interface positions reported by Garoosi et al. The large-scale morphology is accurately reproduced at all three reference times. The root-mean-square distances from the VOF markers to the closest computed interface are respectively $0.0046H$, $0.0108H$, and $0.0152H$ for the lower interface, and $0.0051H$, $0.0124H$, and $0.0209H$ for the upper interface. The third comparison panel is reported at $t=0.975$, while the fourth panel shows the later nonlinear morphology at $t=1.750$.

\begin{figure}[H]
\centering
\includegraphics[width=0.78\textwidth]{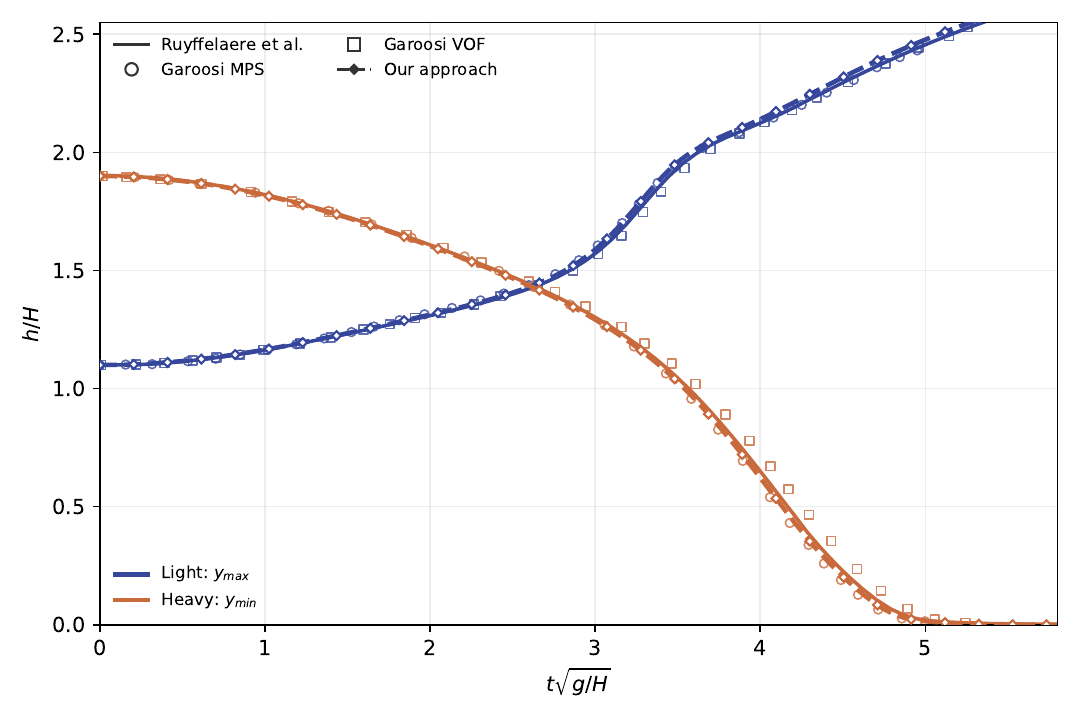}
\caption{Dimensionless temporal evolution of the maximum height of the light phase and the minimum height of the heavy phase. Solid curves denote the results of Ruyffelaere et al.~\cite{Ruyffelaere2026TopologyPFEM}, circles and squares denote the Garoosi MPS and VOF data, and dashed curves with diamonds denote our approach.}
\label{fig:rt_three_phase_extrema}
\end{figure}

For a quantitative comparison, Figure~\ref{fig:rt_three_phase_extrema} reports the maximum height of the light phase and the minimum height of the heavy phase. In accordance with the reference benchmark, %
the dimensionless time is $T=t\sqrt{g/H}$ and the vertical positions are normalized by $H$. Over the interval displayed, the present results have root-mean-square differences of $0.0147H$ and $0.0124H$ relative to the interface-fitted PFEM curves of Ruyffelaere et al.~\cite{Ruyffelaere2026TopologyPFEM} for the light and heavy phases, respectively. The corresponding differences with the Garoosi MPS data are $0.0143H$ and $0.0140H$. The larger difference with the heavy-phase VOF data, $0.0552H$, is consistent with the offset between the MPS and VOF reference series already visible in the benchmark data. Overall, the extrema obtained with the present approach remain within the reference envelope throughout the growth and penetration stages.

\subsection{Rising bubble with moderate deformation}
\label{sec:hysing_case1}

We next consider the first rising-bubble benchmark of Hysing et al.~\cite{hysing2009bubble}. The computation is performed in the rectangular domain $\Omega=[0,1]\times[0,2]$, with an initially circular bubble of radius $0.25$ centered at $(0.5,0.5)$. The flow starts from rest and is driven by gravity, $g=-0.98$. The four walls are treated as free-slip boundaries. The ambient liquid is heavier and more viscous than the gas bubble: in the benchmark units, case 1 uses $\rho_l=1000$, $\rho_g=100$, $\mu_l=10$, $\mu_g=1$, and surface tension $\sigma=24.5$. This parameter set produces a rising bubble that deforms but remains simply connected over the time interval considered.

\begin{figure}[H]
\centering
\includegraphics[width=\textwidth]{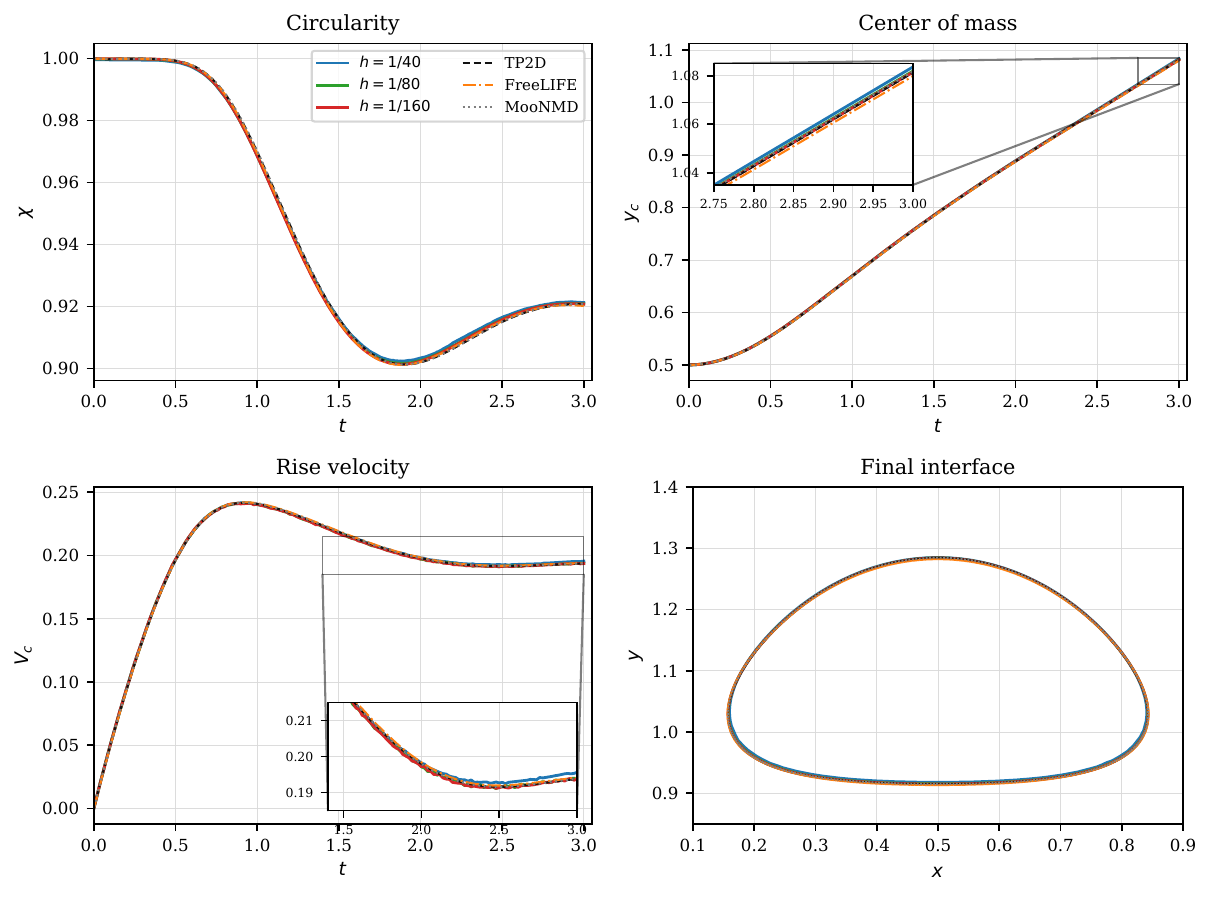}
\caption{Hysing case 1 benchmark. Circularity, vertical center of mass, rise velocity and final interface obtained with the marker-based X-Mesh method are compared with TP2D, FreeLIFE and MooNMD reference results.}
\label{fig:hysing_case1}
\end{figure}

The comparison uses the standard benchmark diagnostics. The circularity is the ratio between the perimeter of a circle with the same area as the bubble and the actual interface length; it is equal to one for a circle and decreases as the bubble deforms. The center of mass is the vertical coordinate of the bubble centroid, and the rise velocity is the bubble-averaged vertical velocity. Figure~\ref{fig:hysing_case1} reports these quantities, together with the final interface, for three mesh sizes and compares them with the TP2D, FreeLIFE and MooNMD reference data for their finest mesh.

For this moderate-deformation case, the computed curves show the expected mesh-convergence trend. The center of mass and the rise velocity are already close to the reference envelope on the intermediate mesh, and the finest computation gives a final bubble position and velocity consistent with the reference values. The circularity minimum is slightly more sensitive to mesh resolution, as it depends directly on the resolved interface length, but the discrepancy decreases with refinement. The final interface comparison is also satisfactory: the cap shape and the lower indentation are recovered without visible loss of symmetry, which indicates that the marker advection and the mesh-conforming surface-tension treatment do not introduce a noticeable geometrical bias in this smooth-interface regime.

\subsection{Rising bubble with strong deformation}
\label{sec:hysing_case2}

The second Hysing benchmark uses the same domain, initial bubble radius, initial position, gravity and boundary conditions, but increases the physical contrast and lowers the surface tension. The material parameters are $\rho_l=1000$, $\rho_g=1$, $\mu_l=10$, $\mu_g=0.1$, and $\sigma=1.96$. The resulting dynamics is substantially more severe than in case 1: the bubble rises faster, develops a flattened cap, and forms long thin filaments near its lower sides. This case is therefore useful for assessing not only the two-phase flow coupling, but also the interaction between the marker representation, the X-Mesh deformation and the explicit topology-control length.

\begin{figure}[H]
\centering
\includegraphics[width=\textwidth]{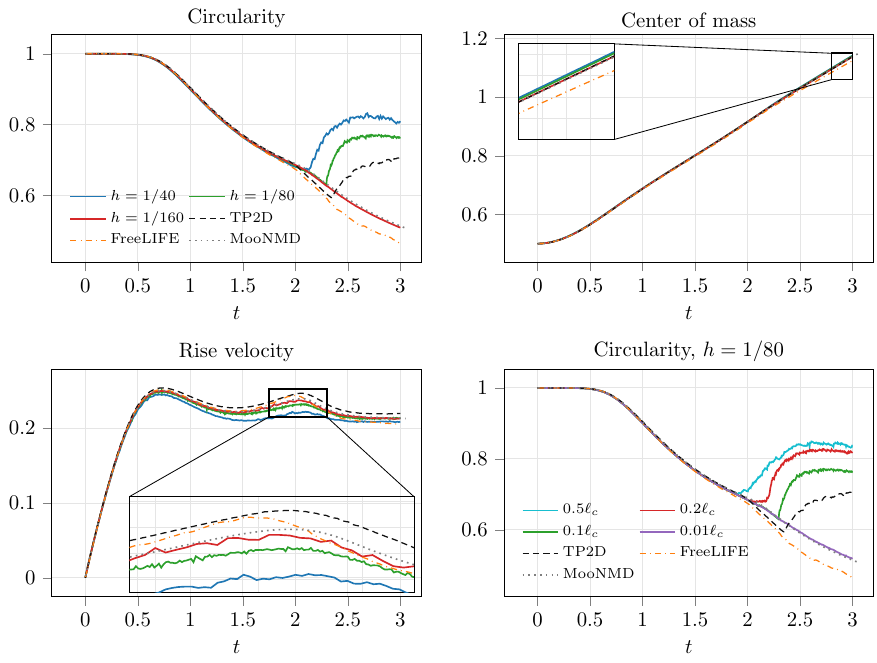}
\caption{Hysing case 2 benchmark. Circularity, vertical center of mass and rise velocity are shown for the mesh sizes $h=1/40$, $h=1/80$ and $h=1/160$ using the topology-control length $0.1\ell_c$, and are compared with the TP2D, FreeLIFE and MooNMD reference data. The bottom-right panel isolates the influence of the topology-control length on the circularity at $h=1/80$.}
\label{fig:hysing_case2}
\end{figure}

Figure~\ref{fig:hysing_case2} shows the circularity, center of mass and rise velocity for several mesh sizes and topology-control thresholds. The early part of the motion is rather insensitive to the cut length: the velocity peak and the initial rise of the bubble are similar for all resolved computations. Differences become visible once the lower skirt starts to stretch. Smaller cut lengths preserve the elongated structures for longer times, which lowers the circularity and modifies the late-time rise velocity. Larger cut lengths remove the finest structures earlier; the corresponding curves remain smoother but no longer represent the same small-scale interface content. This sensitivity is expected, because the topology-control length is an explicit modelling parameter in the present method.

The comparison with the TP2D, FreeLIFE and MooNMD references should therefore be interpreted with this modelling choice in mind. The global rise remains in a comparable range, but the late-time circularity and interface topology depend on how aggressively small ligaments are retained or removed. The finest computation ($h=1/160$) improves the spatial resolution of the cap and of the side filaments, but it does not eliminate the dependence on the topology criterion. In particular, the $0.01\ell_c$ computation keeps thin structures longer than the coarser-cut variants and consequently gives a lower late-time circularity. This is a useful indication that the method is resolving the imposed topological scale, rather than hiding it through numerical diffusion.

This parameter dependence also provides useful modelling flexibility. When topology changes are strongly limited and the thin filaments are allowed to persist, the present results approach those of MooNMD, which is based on a purely ALE interface description. Conversely, increasing the user-selected topology-control length permits the filaments to be cut earlier and shifts the solution towards a regime closer to TP2D. The method can therefore move continuously between connectivity-preserving and breakup-permitting descriptions through a transparent user parameter, providing additional freedom and greater versatility for flows in which the relevant topological scale is problem dependent.

\begin{figure}[H]
\centering
\includegraphics[width=\textwidth]{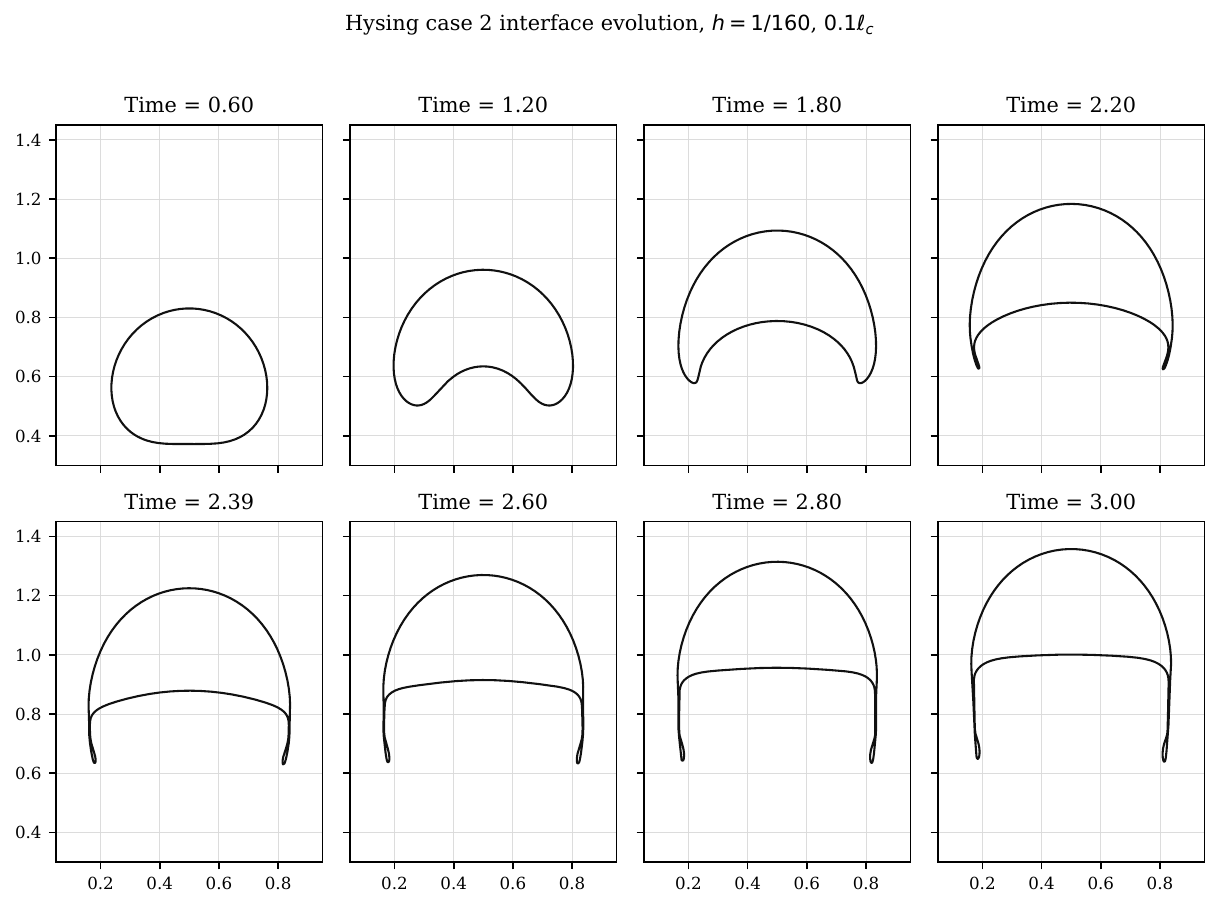}
\caption{Time evolution of the Hysing case 2 interface for the computation with $h=1/160$ and a cut length of $0.1\ell_c$.}
\label{fig:hysing_case2_interface_evolution}
\end{figure}

Figure~\ref{fig:hysing_case2_interface_evolution} shows the interface evolution for the $h=1/160$ computation with cut length $0.1\ell_c$. The sequence makes the physical and numerical difficulty of the test visible. The bubble remains connected throughout the complete interval: it first develops the expected skirted shape and then forms two long, thin lower filaments. Although these filaments continue to stretch up to $t=3$, the topology-control criterion is never triggered in this computation. The late-time geometry is therefore a continuously represented interface, rather than a collection of components produced by an explicit cutting operation.

\begin{figure}[H]
\centering
\includegraphics[width=0.6\textwidth]{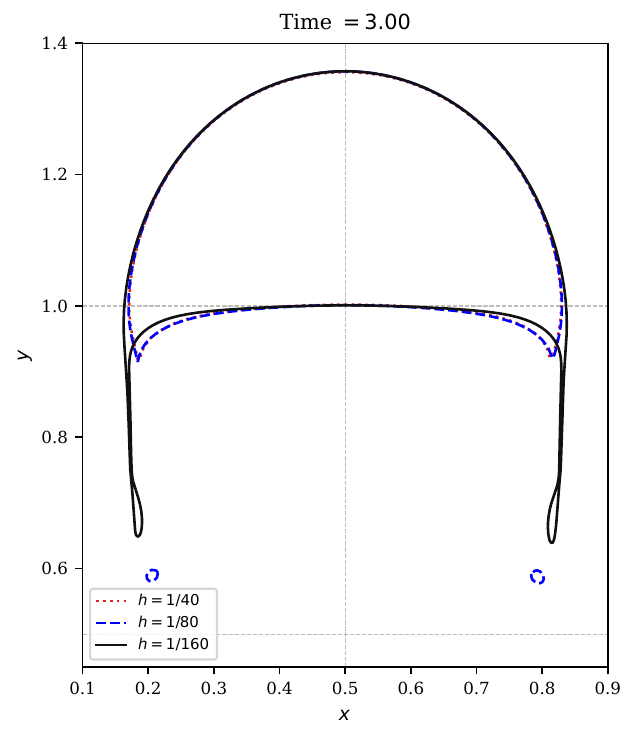}
\caption{Comparison of the Hysing case 2 interface at time $t=3$ for different spatial resolutions: $h=1/40$ (red dotted line), $h=1/80$ (blue dashed line), and $h=1/160$ (black line) for a cut length of $0.1\ell_c$.}
\label{fig:hysing_case2_interface_t3_comparison}
\end{figure}

Figure~\ref{fig:hysing_case2_interface_t3_comparison} compares the interface at $t=3$ for several spatial resolutions at the same relative topology-control length. The upper cap is nearly unchanged, whereas the lower filaments show a strong resolution dependence. For $h=1/40$, the larger absolute cut-off has already removed the thin lower extensions and only the main cap remains. For $h=1/80$, two small detached components are still visible below the cap. At $h=1/160$, no topology change occurs and the two longer filament tips remain connected to the main interface. The comparison thus identifies the lower filaments as the region most sensitive to the absolute feature-size threshold.

\subsection{Triple points and wetting}
\label{sec:wetting_tests}

The preceding test cases involve closed material interfaces. We now consider two configurations in which the fluid--fluid interface terminates on a solid boundary. These tests assess the wall-energy term in Eq.~\eqref{eq:young_wall_energy}, the motion of the resulting triple points, and the recovery of the prescribed Young angle. Only the latest simulation series is reported below.

\subsubsection{Static sessile drops}
\label{sec:static_sessile_drops}

Three initially semicircular drops of radius $0.3$ are placed on the lower wall of the domain $\Omega=[0,2]\times[0,1]$. The two fluids have identical density $\rho=1$ and dynamic viscosity $\mu=0.1$, the fluid--fluid interfacial tension is $\sigma=1$, and gravity is omitted. These are model fluids with equal material properties, rather than a dimensional air--water pair. The solid--fluid surface-energy difference is specified separately through $\gamma_{SL}-\gamma_{SG}=-\sigma\cos\theta_Y$. The characteristic mesh size is $h=0.025$, the time step is $\Delta t=7.8848\times 10^{-4}$, and the computations are advanced to $t=3.1539$. Young angles of $60^\circ$, $90^\circ$, and $120^\circ$ are imposed through the wall-energy term. No topological operation is triggered in any of these calculations. The transient relaxation retains the convective term and uses a fully implicit Navier--Stokes solve at every time step.

\begin{figure}[H]
\centering
\includegraphics[width=\textwidth]{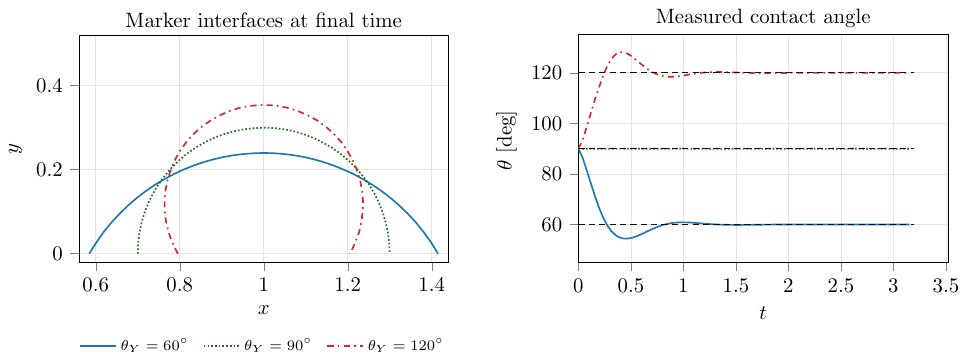}
\caption{Static sessile-drop tests for imposed Young angles of $60^\circ$, $90^\circ$, and $120^\circ$. Left: marker interfaces at $t=3.1539$. Right: contact angles measured from circular fits of the marker chains; dashed lines denote the imposed values.}
\label{fig:triple_contact_angles}
\end{figure}

Figure~\ref{fig:triple_contact_angles} shows the final marker chains and the contact-angle histories. The reported angles are obtained from a circular fit of the marker chain and are $60.0353^\circ$, $90.0070^\circ$, and $119.9400^\circ$, respectively. The corresponding absolute errors remain below $0.060^\circ$. Moreover, the largest peak-to-peak variation over the final five reported values is $5.56\times10^{-4}$ degrees. The three equilibrium geometries therefore recover both acute and obtuse Young angles without introducing a curvature reconstruction or a special force at the triple point.

\subsubsection{Capillary rise}
\label{sec:capillary_rise}

The second wetting test is based on the $\Omega=1$ capillary-rise configuration of Gr\"unding et al.~\cite{Gruending2020CapillaryRise}. By symmetry, only the right half of a channel of half-gap $R=5\,\mathrm{mm}$ and height $40\,\mathrm{mm}$ is simulated. The left boundary is the symmetry plane and the right boundary is the wetting wall; both impose zero normal velocity and leave the tangential velocity unconstrained. The Young angle is $30^\circ$. The liquid and gas densities are $83.1$ and $0.0831\,\mathrm{kg\,m^{-3}}$, respectively, the surface tension is $0.04\,\mathrm{N\,m^{-1}}$, and the gravitational acceleration is $4.17\,\mathrm{m\,s^{-2}}$. The present equilibrium calculation uses $\mu_l=\mu_g=10^{-2}\,\mathrm{Pa\,s}$ to damp late-time parasitic currents. This differs from the gas viscosity and the Navier-slip wall model of the published transient benchmark; consequently, only the viscosity-independent static equilibrium is compared here. The computation uses $h=0.5\,\mathrm{mm}$ and $\Delta t=10^{-4}\,\mathrm{s}$ and is advanced to $t=2.5\,\mathrm{s}$.

\begin{figure}[H]
\centering
\includegraphics[width=\textwidth]{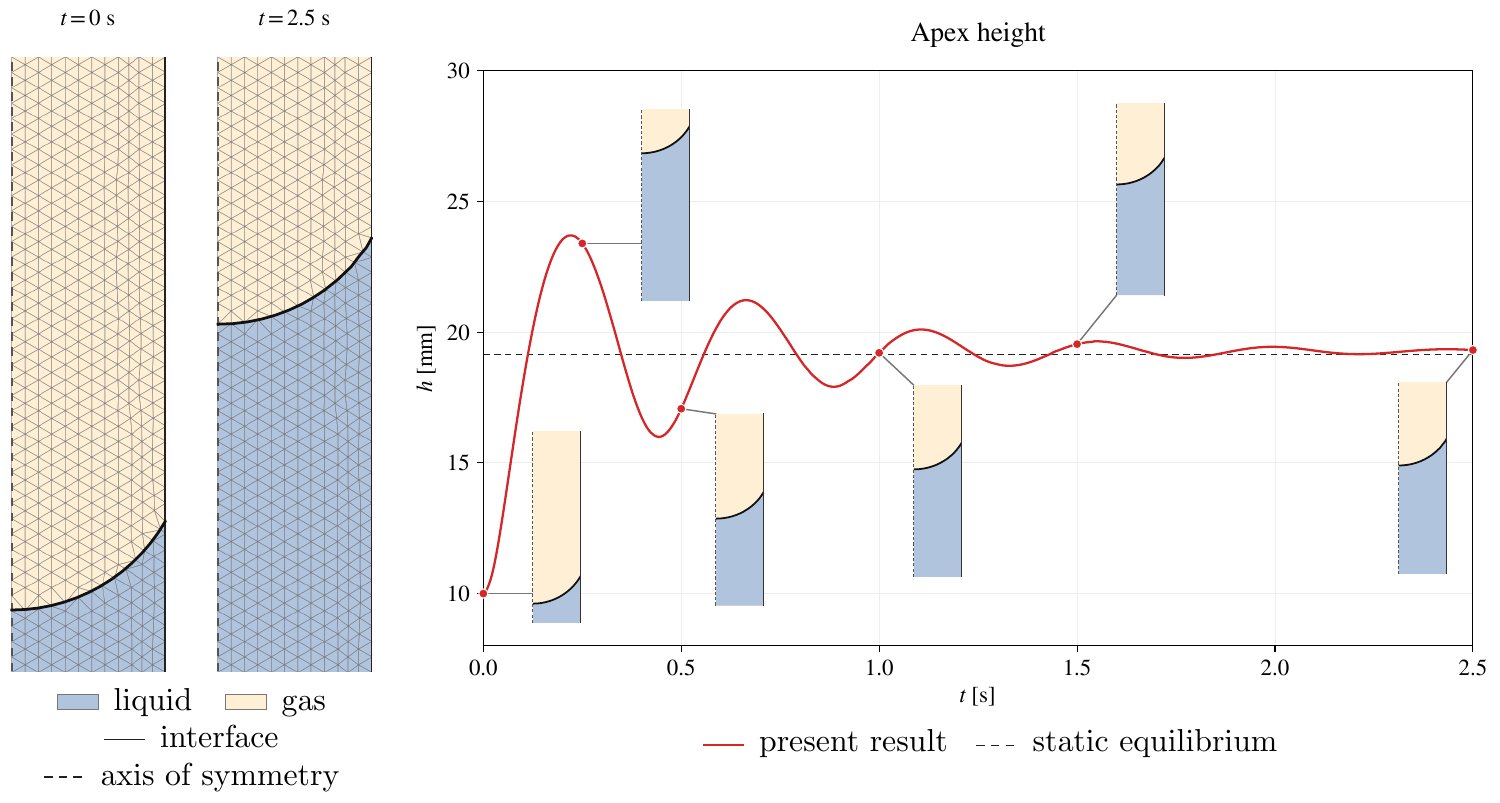}
\caption{Capillary rise for the $\Omega=1$ configuration of Gr\"unding et al.~\cite{Gruending2020CapillaryRise}. Left: fitted meshes at $t=0$ and $t=2.5\,\mathrm{s}$. Right: apex-height history and corrected static equilibrium; dots and connectors associate the six mesh-free snapshots with their times on the curve. All spatial views cover $0\le x\le5$ and $8\le y\le28\,\mathrm{mm}$.}
\label{fig:capillary_rise_mesh_sequence}
\label{fig:triple_capillary_rise}
\end{figure}

Figure~\ref{fig:capillary_rise_mesh_sequence} combines the initial and final fitted meshes with the apex-height history. The six mesh-free snapshots are linked to the curve at their physical times, showing the rise, overshoot and damped relaxation. All spatial views show the computed right half-channel over $8\le y\le28\,\mathrm{mm}$; the dashed vertical boundary is the symmetry plane $x=0$.

The corrected static apex height for this configuration is $h_\infty=19.1538\,\mathrm{mm}$. Figure~\ref{fig:triple_capillary_rise} shows the damped approach toward this value. At the final time, the computed apex height is $19.3174\,\mathrm{mm}$, corresponding to an absolute error of $0.1636\,\mathrm{mm}$ and a relative error of $0.854\%$. Its peak-to-peak variation over the final 1000 time steps is $0.0309\,\mathrm{mm}$. The contact angle measured directly from the phase-separating mesh edges is $30.1284^\circ$ at the final time and varies by only $0.0407^\circ$ over the final $0.5\,\mathrm{s}$. These results complement the sessile-drop tests by showing that the same membrane and wall-energy formulation sustains a hydrostatic meniscus while retaining the prescribed triple-point condition.

\begingroup
\subsection{Three-fluid lenses with free triple junctions}
\label{sec:liquid_lenses}

We consider a capillary-relaxation benchmark involving three immiscible fluids
and two free triple junctions. Unlike the wall-contact examples, all three
branches incident to each junction are fluid interfaces. Their equilibrium
angles follow from the balance of the three interfacial tensions, rather than
from a prescribed wall contact angle. Liquid lenses provide a direct test of
this balance~\cite{Semprebon2016Ternary}.

The domain is the unit square. Fluid~0 occupies the lower region, fluid~1 the
upper region, and fluid~2 forms a lens intersecting their interface. Gravity
is zero, the initial velocity is zero, and impermeable free-slip conditions
are used on all four walls. The solid surface energies are equal, so that
the outer interface meets the vertical walls at $90^\circ$. We use
$\rho_0=\rho_1=\rho_2=1$ and $\mu_0=\mu_1=\mu_2=0.1$ in consistent simulation
units. Keeping these properties fixed isolates the effect of the interfacial
tensions. Density and viscosity influence the transient response but do not
enter the gravity-free static angle balance.

All three computations start from the same non-equilibrium interface.

Let $\theta_i$ denote the interior angle occupied by fluid~$i$ at a junction,
and let $\boldsymbol{t}_{ij}$ point from that junction into interface $ij$.
The Neumann condition and the angle convention are
\begin{equation}
\label{eq:lens_neumann}
 \sigma_{01}\boldsymbol{t}_{01}
 +\sigma_{02}\boldsymbol{t}_{02}
 +\sigma_{12}\boldsymbol{t}_{12}=\boldsymbol{0},
 \qquad \theta_0+\theta_1+\theta_2=2\pi.
\end{equation}
The tension ratios and equilibrium angles of all three cases in
Table~\ref{tab:lens_cases} satisfy the relations in Eq.~\eqref{eq:lens_neumann}.

\begin{table}[htbp]
\centering
\begin{tabular}{c|ccc|ccc}
Case & $\sigma_{01}$ & $\sigma_{02}$ & $\sigma_{12}$
     & $\theta_0$ & $\theta_1$ & $\theta_2$\\ \hline
A & $1$ & $1$ & $1$ & $120^\circ$ & $120^\circ$ & $120^\circ$\\
B & $1$ & $\sqrt{3}/2$ & $1/2$ & $150^\circ$ & $120^\circ$ & $90^\circ$\\
C & $1$ & $1/\sqrt{3}$ & $1/\sqrt{3}$ & $150^\circ$ & $150^\circ$ & $60^\circ$
\end{tabular}
\caption{Interfacial tension ratios and exact equilibrium fluid-sector angles.
Fluid~2 is the lens. Angles are not the interior angles of the force triangle.}
\label{tab:lens_cases}
\end{table}

The lens angle $\theta_2$ is measured at both triple junctions by fitting
circles independently to the upper and lower marker chains. The angle is
then evaluated between the tangents of the fitted circles at each junction.

\subsubsection*{Computed results}
The three runs below use $h=0.025$ and $\Delta t=6.308\times10^{-4}$.
Table~\ref{tab:lens_measured} gives the actual final times and measured lens
angles at the left/right junctions. The area column is the maximum relative
drift among all three marker regions over the complete run.
Figure~\ref{fig:lens_profiles_mesh} shows the terminal profiles and the
computational mesh near the right triple junction.

\begin{table}[htbp]
\centering\small
\begin{tabular}{c|c|c|c|c}
Case & $t_f$ & fitted $\theta_2$ (deg.) & area drift (\%) & $\|u\|_\infty$\\\hline
A & 2.8385 & 120.222/120.232 & 0.683 & 4.73e-03 \\
B & 2.8385 & 90.073/90.072 & 0.506 & 2.93e-04 \\
C & 2.8385 & 60.058/60.058 & 0.080 & 1.16e-04 \\
\end{tabular}
\caption{Measured terminal states. Lens angles at the left/right junctions
are obtained from circular fits of the two full marker arcs.}
\label{tab:lens_measured}
\end{table}

\begin{figure}[H]
\centering
\includegraphics[width=\textwidth]{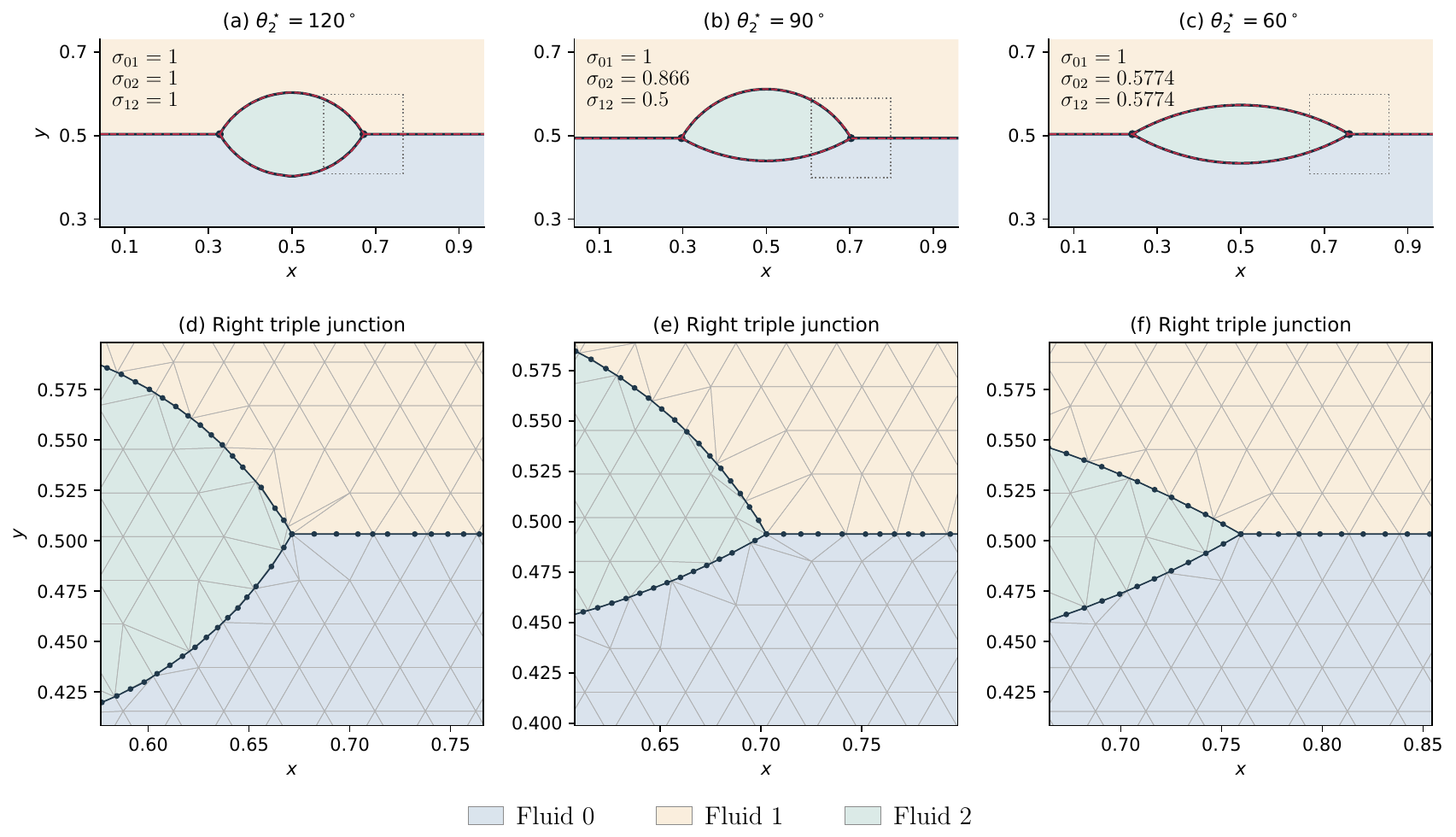}
\caption{Three-fluid lens calculations at $t=2.8385$ for target lens angles
of $120^\circ$, $90^\circ$ and $60^\circ$ (columns, left to right).
Top: terminal marker profiles (solid) and analytical reference profiles
(red dashed). Grey boxes locate the regions enlarged below.
Bottom: computational mesh and marker chains near the right triple junction
at the same terminal time.}
\label{fig:lens_profiles_mesh}
\end{figure}

\par
\endgroup

\section{Conclusion}

This work introduced a marker-based X-Mesh formulation for sharp-interface $N$-phase flow simulations. The central idea is to separate the geometrical representation of the interface from the computational mesh, while still deforming the mesh locally so that selected finite element edges conform to the tracked front. In contrast with a purely level-set description, the interface is not inferred from nodal signs. This avoids the ambiguous situation in which all the vertices of an element lie on the zero level set: such an element has no well-defined sign and may be interpreted simultaneously as belonging to either phase. With the present marker representation, the interface geometry is explicit, and the phase of the elements is reconstructed from the marker topology. The phase description therefore remains well defined even when the mesh contains highly degenerated elements aligned with the interface.

The fitted interface is coupled to a pressure space that is continuous within each phase and to a variational membrane representation of surface tension. The pressure traces on the two sides of the interface are independent, so the capillary pressure jump is obtained as part of the solution without prescribing a discrete curvature. At a solid wall, the same energy-based construction incorporates Young's law and permits fluid--fluid--solid triple points to move tangentially along the boundary. The sessile-drop and capillary-rise tests recover the prescribed Young angles for both acute and obtuse wetting configurations and sustain the corresponding static meniscus. The three-phase Rayleigh--Taylor and liquid-lens examples demonstrate phase-partition reconstruction and free triple junctions inside the domain.

This decoupling is also what allows the method to retain interfacial filaments that are thinner than the characteristic size of the background mesh. The marker chain carries the geometry, while the X-Mesh deformation creates the local degeneracies needed to embed that geometry in a conforming finite element mesh without remeshing. Part of the constraint imposed by the computational mesh is further alleviated by the Winslow-based bulk deformation: by redistributing nodes toward the tracked interface, the method locally increases the number of degrees of freedom available to represent interfacial features while preserving the original mesh connectivity. The numerical examples illustrate this behavior in both prescribed-advection and coupled $N$-phase flow configurations: thin structures can be transported sharply, topology changes can be introduced through an explicit feature-size criterion, and standard rising-bubble diagnostics remain consistent with established benchmark data in the smooth-interface regime.

The same property also defines an important limitation of the approach. Since the interface resolution is no longer tied to the computational mesh size, the method can generate or preserve structures that are much finer than the scale at which the flow solution is accurately resolved. In that regime, the marker geometry may remain sharp while the velocity, pressure and surface-tension forces are under-resolved on the surrounding mesh. This scale separation can lead to oscillations, a known difficulty of marker and front-tracking methods when the front resolution is not kept compatible with the fluid discretization \cite{popinet_fronttracking}. In practice, the marker spacing, the mesh size and the topology-control length must therefore be chosen together. The cut-off length used in the present work is not only a numerical safeguard; it is a modelling parameter that prevents the computation from retaining geometrical features whose hydrodynamic effect cannot be represented reliably on the available mesh.

Future work will focus on extending the same ideas to three-dimensional marker surfaces and to more complex topology changes, where preserving an unambiguous phase description without remeshing remains the main motivation for the marker-based X-Mesh approach. A central challenge in this extension will be to maintain the validity of the deformed volume mesh, and in particular to guarantee that the interface-conforming deformation does not produce inverted elements.

\section*{CRediT authorship contribution statement}
\textbf{Antoine Quiriny:} Conceptualization, Methodology, Software, Investigation, Writing -- original draft.

\textbf{Jonathan Lambrechts:} Conceptualization, Software, Supervision, Writing -- review \& editing.

\textbf{Michel Henry:} Software, Writing -- review \& editing.

\textbf{Nicolas Moës:} Supervision, Writing -- review \& editing, Funding acquisition.

\textbf{Jean-François Remacle:} Conceptualization, Supervision, Writing -- review \& editing, Funding acquisition.

\section*{Acknowledgements}
The authors acknowledge support from the European Research Council (ERC) through the X-MESH project (grant agreement No.~101071255), funded under the European Union's Horizon Europe research and innovation programme.

\section*{Declaration of competing interest}
The authors declare that they have no competing financial interests or personal relationships that could have influenced the work reported in this article.

\section*{Data and code availability}
The simulation data and source code are not publicly available at the time of submission.

\section*{Declaration of generative AI and AI-assisted technologies in the manuscript preparation process}
The authors used OpenAI Codex to assist with drafting and revising manuscript text and preparing the submission materials. The authors reviewed and edited the AI-assisted content and take full responsibility for the content of the article.

\bibliographystyle{plain}
\bibliography{main}

\end{document}